\documentclass{aa}  

\usepackage{natbib,twoopt}
\usepackage[breaklinks=true]{hyperref} %% to avoid \citeads line fills
\bibpunct{(}{)}{;}{a}{}{,}             %% natbib format for A&A and ApJ
\makeatletter
  \newcommandtwoopt{\citeads}[3][][]{\href{http://adsabs.harvard.edu/abs/#3}%
    {\def\hyper@linkstart##1##2{}%
     \let\hyper@linkend\@empty\citealp[#1][#2]{#3}}}
  \newcommandtwoopt{\citepads}[3][][]{\href{http://adsabs.harvard.edu/abs/#3}%
    {\def\hyper@linkstart##1##2{}%
     \let\hyper@linkend\@empty\citep[#1][#2]{#3}}}
  \newcommandtwoopt{\citetads}[3][][]{\href{http://adsabs.harvard.edu/abs/#3}%
    {\def\hyper@linkstart##1##2{}%
     \let\hyper@linkend\@empty\citet[#1][#2]{#3}}}
  \newcommandtwoopt{\citeyearads}[3][][]%
    {\href{http://adsabs.harvard.edu/abs/#3}
    {\def\hyper@linkstart##1##2{}%
     \let\hyper@linkend\@empty\citeyear[#1][#2]{#3}}}
\makeatother

\usepackage{color}
\hypersetup{colorlinks=true,linkcolor=blue,citecolor=blue,urlcolor=blue}

\usepackage[breaklinks=true]{hyperref}

\usepackage{graphicx}
\usepackage{txfonts}
\usepackage{amsmath}	% Advanced maths commands
\usepackage{caption}
\usepackage{subcaption}
\usepackage{hhline}
\usepackage{multirow}
\usepackage[flushleft]{threeparttable}
\usepackage{lscape}
\usepackage{comment}
\usepackage{url}

\def\ms{\,m\,s$^{-1}$}         %m.s -1
\def\Mearth{\hbox{$\mathrm{M}_{\oplus}$}}

\def\degr{\hbox{$^\circ$}}

\renewcommand*{\vec}[1]{\mathbf{#1}}
\newcommand*{\mat}[1]{\mathbf{#1}}

\begin{document}

  \title{An improved correction of radial-velocity systematic for the SOPHIE spectrograph}

   \author{S.~Grouffal\inst{1}
          \and
          A.~Santerne\inst{1}
          \and
          N.C.~Hara\inst{1,4}
          \and
          I.~Boisse\inst{1}
          \and
          S.~Coez\inst{2}
          \and 
          N.~Heidari\inst{3}
          \and
          S.~Sulis\inst{1}
          }

   \institute{Aix Marseille Univ, CNRS, CNES, Institut Origines, LAM, Marseille, France
         \and
             ENS Paris-Saclay, Gif sur Yvette, France
             \and
             Institut d’astrophysique de Paris, UMR 7095 CNRS université pierre et marie curie, 98 bis, boulevard Arago, 75014, Paris
             \and
             Observatoire Astronomique de l’Université de Genève, 51 Chemin de Pegasi b, 1290 Versoix, Switzerland
             }

   \date{Received 15 December 2023; accepted 24 April 2024}

% \abstract{}{}{}{}{} 
% 5 {} token are mandatory
 
  \abstract
   {High precision spectrographs might exhibit temporal variations of their reference velocity or nightly zero point (NZP). One way to monitor the NZP is to measure bright stars, which are assumed to have an intrinsic radial velocity variation much smaller than the instrument's precision. Their variations, primarily assumed to be instrumental, are then smoothed into a reference radial velocity time series (master constant), which is subtracted from the observed targets. While this method is effective in most cases, it does not fully propagate the uncertainty arising from NZP variations.
We present a new method to correct for NZP variations in radial-velocity time series. This method uses Gaussian Processes based on ancillary information to model these systematic effects. Moreover, it enables us to propagate the uncertainties of this correction into the overall error budget. Another advantage of this approach is that it relies on ancillary data collected simultaneously with the spectra rather than solely on dedicated observations of constant stars.  
We applied this method to the SOPHIE spectrograph at the Haute-Provence Observatory using a few instrument's housekeeping data, such as the internal pressure and temperature variations. Our results demonstrate that this method effectively models the red noise of constant stars, even with a limited amount of housekeeping data, while preserving the signals of exoplanets. Using both simulations with mock planets and real data, we found that this method significantly improves the false-alarm probability of detections by several orders of magnitude. Additionally, by simulating numerous planetary signals, we were able to detect up to 10 $\%$ more planets with small amplitude radial velocity signals. We used this new correction to reanalysed the planetary system around HD158259 and improved the detection of the outermost planets. %result
We propose this technique as a complementary approach to the classical master-constant correction of the instrumental red noise. We also suggest decreasing the observing cadence of the constant stars to optimise telescope time for scientific targets. }

    
    \maketitle
%
%-------------------------------------------------------------------

\section{Introduction}

The radial-velocity (RV) technique is one of the most prolific to detect and characterise exoplanets. Since the discovery of 51~Peg~b \citep{1995Natur.378..355M}, a total of 1079 planets \footnote{As listed on The Exoplanet Encyclopedia (\url{http://exoplanet.eu/}) accessed November 23, 2023} have been discovered using the RV technique. This technique remains the most efficient to measure the mass of exoplanets, which is essential to understand their composition.

However, instruments suffer from systematics. It concerns mainly instruments that are not in vacuum chambers like the SOPHIE spectrograph \citep{2008SPIE.7014E..0JP, 2009A&A...505..853B} mounted on the 1.93m telescope of the Haute-Provence Observatory\footnote{\url{http://www.obs-hp.fr/guide/sophie/sophie-eng.shtml}}. Despite being stabilised in pressure and temperature, the SOPHIE spectrograph is still sensitive to small environmental variations that are not perfectly monitored by calibrations. This leads to a long-term variation of the spectrograph's nightly-zero-point (hereafter NZP), as presented by \citet{2015A&A...581A..38C}. 

To correct for these instrumental variations, a sliding median is used to build an estimate of the NZP based on the RV variations of a few bright and assumed constant stars. This NZP model, known as the master constant, is subtracted from the RV time series measured with SOPHIE \citep[\citet{Heidari}]{2015A&A...581A..38C}. This method has been shown to significantly decreases the RV root mean square (RMS) of the star HD185144 from 5.4~\ms\, to 1.6~\ms\, \citep{2015A&A...581A..38C}, enabling the detection of low-mass planets \citep{2018A&A...618A.103H,2019A&A...625A..18H, 2019A&A...625A..17D,2020A&A...636L...6H}.  
However, by subtracting the master constant from the RV time series before their analysis, the error budget is not fully propagated. This can potentially affect the false-alarm probability (FAP) of detections and introduce uncertainties in the estimation of planetary parameters.

Gaussian processes (hereafter GPs) are now increasingly used in many domains of astrophysics, especially to model stochastic processes in both photometry and RV time series \citep[e.g., see review by][]{2022arXiv220908940A}. They provide a flexible and mathematically simple approach to model stochastic processes. GPs could be an interesting alternative for correcting NZP variations in RV spectrographs like SOPHIE.

In this work, we propose a method that utilises GPs based on ancillary information to improve the correction of NZP variations in RV time series. This method is inspired by the one commonly used to correct for space-based photometry in transit detection \citep{2012MNRAS.419.2683G, 2016MNRAS.459.2408A} and transmission spectroscopy \citep{2015MNRAS.451..680E, 2017Natur.548...58E}. Using GPs, we can handle the correlation between the RVs and ancillary information without choosing any deterministic functional form. We applied this method to SOPHIE data using as ancillary information the housekeeping data, such as pressure and temperature, from the spectrograph measured every 6 minutes. 

This paper is organised as follows. Section \ref{section2} presents the different housekeeping data and their potential impact on the RV time series. Section \ref{section3} describes the GP model and the proposed method. In section \ref{section4}, we apply the method on SOPHIE constant stars and tested it on mock planets, highlighting its improvement in detecting low-mass planets. Section \ref{section5} demonstrates the application of the method on the system HD 158259. Finally, in Section \ref{section6}, we present our conclusions, discuss the impact on observing strategies and planet detection with SOPHIE, and consider possible avenues for further improvement.

\section{Systematic effects in radial-velocity spectrographs}
\label{section2}

\begin{figure}
   \centering
   \includegraphics[width=\columnwidth]{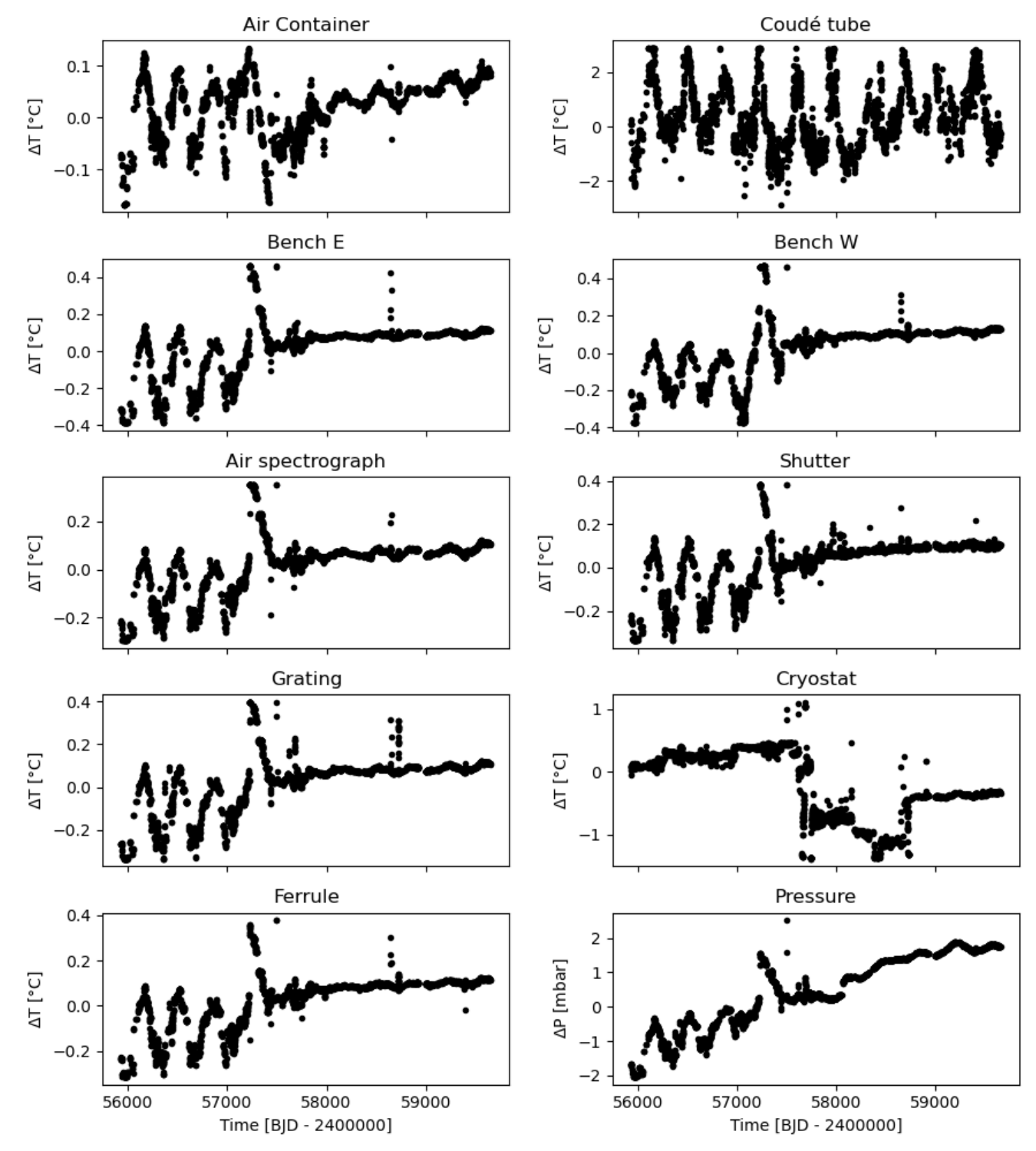}
      \caption{Housekeeping data variations monitored at different locations of the SOPHIE instrument with a cadence of 6 minutes. The pressure corresponds to the internal pressure inside the Nitrogen-filled tank where the dispersal elements are located. The temperature is monitored in the air of the container, the air in the spectrograph, the shutter, the east and west granite benches, the grating, the cryostat cover, the ferrule, and the Coudé tube (path of the fibres from the telescope to the spectrograph). See \citet{2008SPIE.7014E..0JP} and the Haute-Provence Observatory website for more details. Near BJD=2457700, the SOPHIE thermal regulation was upgraded, improving the temperature stability inside the spectrograph.}
         \label{housekeeping_variables}
   \end{figure}

The RV time series of a star is the addition of the astrophysical signal, some correlated noise from systematics, and uncorrelated white noise. These systematics might result from variations in the spectral indices of the air inside the spectrograph, especially if it is not a vacuum tank. Fluctuations in temperature and pressure are responsible for the change of spectral indices. For instance, a small variation of 1~mbar in pressure might induces a spectral shift up to 300 \ms, and a temperature variation of 1 K might causes a shift up to 90 \ms\, \citep{2010EAS....41...27E}. To limit these variations, the SOPHIE spectrograph is located in a pressure and temperature-controlled chamber. Great efforts have already been made by the observatory staff to limit these variations. However, some fluctuations of the internal pressure and temperature measured in the spectrograph environment are still observed (see figure \ref{housekeeping_variables} and online table \ref{table_housekeeping}) and are expected to cause RV systematics. 

\begin{figure}
     \centering
     \includegraphics[width=\columnwidth]{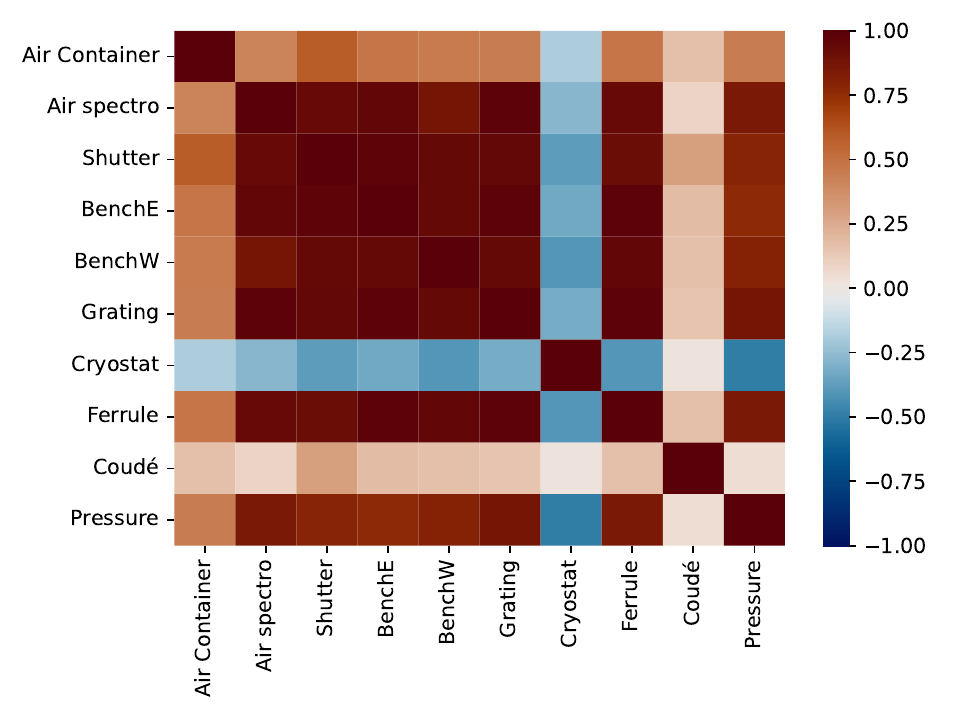}

    \caption{Heat map of Pearson coefficients to evaluate the linear correlation between the 10 housekeeping variables available for the SOPHIE instrument.}
    \label{heat_map}
\end{figure}

Figure \ref{housekeeping_variables} shows that several of these housekeeping data exhibit similar patterns, suggesting they are correlated. To evaluate this correlation, the heat map in Fig. \ref{heat_map} shows the Pearson coefficient between all the housekeeping variables presented in Figure \ref{housekeeping_variables}. The containers, the ferrule, the east and west bench, and the grating are correlated with each other with a Pearson coefficient above 0.5 and up to 0.98. Only the temperature at the Coudé tube is not correlated with the other parameters. The cryostat also shows a different behaviour with a strong negative linear correlation with the other parameters. As a consequence, we only consider as housekeeping data (i) the internal pressure and the temperatures of the (ii) air in the container, (iii) the Coudé tube, and (iv) the cryostat.
   
\section{Gaussian Processes to correct for nightly-zero-point variations}
\label{section3}

\subsection{Gaussian Process model}
\label{section_GP_model}
A GP is a type of stochastic process $y(x)$ such that, for any finite collection of input $x_1, ..., x_N$, the distribution of vector ($y_{x_1}, ..., y_{x_N}$) is a multivariate Gaussian distribution \citep{Rasmussen}. We can write the joint probability distribution $p(\mathbf{y})$ over the finite sample (of dimension $N$) of data $\mathbf{y} = \{y_i\}_{i=1,...,N}$ from the GP as:

\begin{equation}
    p(\mathbf{y}) = \mathcal{N}(\mathbf{m},\mathbf{C}),
\end{equation}

where $\vec m$ is the mean vector and represents the other deterministic effects, such as the proper motion of the star or constant offset of the instrument, and $\mathbf{C}$ is the covariance matrix such as $\vec C_{ij} = k(|t_i - t_j|)$ with $k(|t_i - t_j|)$ commonly called the kernel. 
\newline

Suppose we have the radial velocity time series of a star $\vec y_0(\vec t_0)$ sampled at times $t_0 = (t_0^1,....,t_0^{N_0})$, such that at each time we have available housekeeping data $(\vec x(t_0^i))$ like pressure and temperature. This star is the one around which we are looking for exoplanets. We also have time series of $p$ constant stars $\vec y_i(\vec t_i)$, $i=1,...,p$. For each $i$, $\vec y_i(\vec t_i)$ is the time series of RV of constant star $i$, sampled at times $t_i = (t_i^1,....,t_i^{N_i})$. We assume that all the stars are affected by the same nightly zero point shift, $z(t)$. This is modelled as a Gaussian Process whose mean is supposed to be zero and whose kernel $k$ is defined over variables $\vec x$ and depends on hyperparameters $\vec \Phi$. In this work, we ignore the uncertainties on the input variables such as pressure and temperature.
The covariance between $z$ sampled at time $t_i$ and $t_j$ is $k(\vec x_i(t_i), \vec x_j(t_j); \vec \Phi)$. The covariance of $z$ sampled at times $\vec t_i$ and $\vec t_j$ is denoted by $\mat V^{ij}(\vec \Phi)$.  

We further assume that their measurements are affected by photon noise. The photon noise affecting star $i$ measured at time $t_i^j$ is denoted by  $\epsilon_i(t_i^j)$. The vectors $\epsilon_i$ are assumed to be Gaussian, with a vanishing mean and a covariance $\mat \Sigma_{i}$. Finally, each star is assumed to have a deterministic mean function depending on variables $\vec \theta_i$, $m_i(t;\theta_i)$.

\begin{align}
    y_0(t) &= m_0(t) +  z(t) +\epsilon_0(t) \\
    y_1(t) &= m_1(t) +  z(t) +\epsilon_1(t) \\ 
    &\vdots  \\
    y_p(t) &= m_p(t) +  z(t) +\epsilon_p(t)  
\end{align}

If we concatenate the column vectors $(\vec y_i(\vec t_i))$ with $i=0,...,p$ into a single column vector of size $\sum_{i=0}^p N_i $, $\vec Y$. Then, the model of the data is fully described by the mean and covariance of $\vec Y$. Its mean $\vec \mu(\vec \theta_0,...\vec \theta_p)$ is the concatenation of vectors $(\vec m_i(\vec t_i))$ $i=0..p$, and its covariance $\mat C$ is a block-matrix such that its $i,j$-th element is a $N_i \times N_j$ matrix:

\begin{equation}
    \mat C_{i,j}(\vec \Phi) = \mat V^{ij} + \delta_{ij} \mat \Sigma_{i}
\end{equation}

where $\delta_{ij} $ is the Kronecker symbol. Or equivalently

\begin{equation}
\mat C(\vec \Phi) = 
  \begin{pmatrix}
\mat V^{00}(\vec \Phi) + \mat \Sigma_0 & \mat V^{01}(\vec \Phi) & \cdots & \mat V^{0p}(\vec \Phi)  \\  \mat V^{01}(\vec \Phi)^T & \mat V^{11}(\vec \Phi) + \mat \Sigma_1 & \cdots & \mat V^{1p}(\vec \Phi) \\
\vdots & \vdots  & \ddots & \vdots \\  \mat V^{0p}(\vec \Phi)^T & \mat V^{p1}(\vec \Phi)  & \cdots & \mat V^{pp}(\vec \Phi) + \mat \Sigma_p
\end{pmatrix}  
\end{equation}

We rewrite $\mat C$ as

\begin{equation}
\mat C(\vec \Phi) = 
\begin{pmatrix}
\mat V^{00}(\vec \Phi) + \mat \Sigma_0 & {\mat V^{0,1:}(\vec \Phi)} \\
{\mat V^{0,1:}(\vec \Phi)}^T & \mat V^{1:,1:}(\vec \Phi) + \mat \Sigma_{1:}
\end{pmatrix}
\end{equation}

where $\mat \Sigma_{1:}$ is a block diagonal matrix with matrices $\mat \Sigma_i$ on the diagonal. ${\mat V^{0,1:}(\vec \Phi)} $ is a matrix of size $N_0 \times \sum_{i=1}^p N_i$ which can be described as $N$ blocks each of size $N_0 \times N_i$ made of $ \mat V^{0i}(\vec \Phi)$. Finally, $\mat V^{1:,1:}(\vec \Phi)$ is a  $\sum_{i=1}^p N_i \times \sum_{i=1}^p N_i$ matrix. 
\newline

The variables $\mu(\vec \theta_0,...,\vec \theta_p)$ which is denoted $\mu(\vec \theta)$ for simplicity, and $\vec C(\vec \phi)$ fully describes the likelihood of the concatenated vector of all available data $\vec Y$, $p(\vec Y \mid \vec \theta, \vec \Phi)$:

\begin{equation}
    p(\vec Y \mid \vec \theta, \vec \Phi) = \mathcal{N}(\vec \mu, \vec C)
    \label{likelihood}
\end{equation}

In principle, this likelihood can be used directly to infer the parameters $\vec \theta$ and $\vec \Phi$. However, this requires the inversion of the full matrix $\vec C$, which might be costly.

To infer the GP's hyperparameters of the kernel, we proceed as follows: we first interpolate linearly the values of each housekeeping variable at the times of the observation after removing outliers at more than 3-$\sigma$ from the median. The housekeeping variables are measured every 6 minutes, and we choose the closest value to the time of the star's observation.
Each variable is normalised at a zero mean and divided by its standard deviation. Finally, we compute the posterior distribution of the parameters of our model, $\Theta$ and $\Phi$, based on the data of the constant stars.

Using the Bayes theorem, we write the posterior distribution as:

\begin{equation}
    p(\vec \Theta,\vec \Phi|\vec Y) \propto p(\vec Y|\vec \Theta,\vec \Phi)p(\vec \Theta,\vec \Phi)
\end{equation}
with $p(\vec Y|\vec \Theta,\vec \Phi)$ the likelihood defined in equation \ref{likelihood}, and $p(\vec \Theta,\vec \Phi)$ the prior on the parameters. 

We can derive the posterior distribution for each hyperparameter and planetary parameter using a Bayesian method. We use the Markov Chain Monte Carlo (MCMC) method as implemented into the \texttt{emcee} package \citep{2013PASP..125..306F} to explore the joint posterior distribution of the parameters. The convergence of the MCMC was tested using the integrated autocorrelation time that quantifies the Monte Carlo error and the efficiency of the MCMC \citep{2010CAMCS...5...65G}. We checked that all the chains are longer than 50 times the integrated autocorrelation time.

\subsection{Choice of the kernel}
\label{subsection_kernels}

The covariance kernel $k(\vec x_i, \vec x_j; \vec \Phi)$, quantifies the self-similarity of the GP. In classical least-square regression, a simple and diagonal covariance matrix is assumed, and the variance associated with each observation is known. However, this assumption does not generally hold, especially when considering instrumental systematics. The covariance kernel addresses this issue, allowing for more complex and flexible covariance models.
The choice of the kernel is a fundamental part of GP regressions as it encodes the structure of the random variations of the process. 
In our study, we tested two different kernels: the Square-exponential kernel and the Matérn 3/2 kernel \citep{Rasmussen}. Additionally, following the approach of \citet{2014MNRAS.445.3401G}, we tested two methods for combining the kernels of each housekeeping variable. The first method involves using a kernel with the product of each individual kernel, which captures the dependencies between non-independent housekeeping variables:

\begin{itemize}

\item Product of Square exponential (SE) multivariate kernel:
\begin{equation}
    k_{PSE}(\vec x_i,\vec x_j, \Phi) = c^2\prod^M_{m=1}  \exp{\left[-\frac{(x_{i,m} - x_{j,m})^2}{l_m^2}\right]}.
    \label{multSE}
\end{equation}

\item Product of Matérn-3/2 (M3) multivariate kernel:
\begin{equation}
    k_{PM3}(\vec x_i,\vec x_j, \Phi) = c^2 \prod^M_{m=1}  \left(1 + \sqrt{3}\frac{x_{i,m} - x_{j,m}}{l_m}\right)e^{-\sqrt{3}\frac{(x_{i,m} - x_{j,m})}{l_m}}.
    \label{multM3}
\end{equation}

According to the strong linear correlations shown in figure \ref{heat_map}, this seems to be the preferred approach as the housekeeping parameters are likely caused by the same physical process and hence are dependent on each other.  
The other solution we tested is a sum of the various kernels representative of a set of independent variables: 

\item A sum of SE multivariate kernel:
\begin{equation}
    k_{SSE}(\vec x_i,\vec x_j, \Phi) = \sum^M_{m=1} c_m^2 \exp{\left[-\frac{(x_{i,m} - x_{j,m})^2}{l_m^2}\right]}.
    \label{addSE}
\end{equation}

\item Sum of M3 multivariate kernel:
\begin{equation}
    k_{SM3}(\vec x_i,\vec x_j, \Phi) = \sum^M_{m=1} c_m^2 \left(1 + \sqrt{3}\frac{x_{i,m} - x_{j,m}}{l_m}\right)e^{-\sqrt{3}\frac{(x_{i,m} - x_{j,m})}{l_m}}.
    \label{addM3}
\end{equation}

\end{itemize}

Here, $c$ is the height scale hyperparameter, and $l$ is the inverse length-scale hyperparameter controlling the correlation length. Kernels with few hyperparameters like square-exponential or Matérn 3/2 are preferred over other kernels like the rational quadratic \citep[eq. 4.19 in][]{Rasmussen} to avoid the GPs being too flexible. As the impact of each variation on the RV time series is unknown, we need to test the two approaches.

Time-dependent kernels were also initially considered but were too flexible and significantly affected the planetary signals. In the worst cases, the planetary signal completely vanishes regardless of the functional forms of the kernel.

\subsection{Two NZP corrections: a GP master constant and noise model}

We suggest two approaches to correct NZP variations based on Gaussian processes. 
Following equations 2.23 and 2.24 from \citet{Rasmussen}, the mean and covariance of the distribution of $\vec y_i (t_0)$ conditioned on  $\vec y_i (t_i)$, $i=1,..p$, denoted by $\vec m_0'$ and $\mat V_0'$ are

\begin{equation}
\vec m_0' =  \vec m_0(\vec t_0; \vec \theta_0) +  {\mat V^{0,1:}(\vec \Phi)} \left( \mat V^{1:,1:}(\vec \Phi) + \mat \Sigma_{1:}  \right)^{-1} \vec R_{1:}
\label{eq:mu0}
\end{equation}

\begin{equation}
\mat V_0' =  \mat V^{00}(\vec \Phi) + \mat \Sigma_0 - {\mat V^{0,1:}(\vec \Phi)} \left( \mat V^{1:,1:}(\vec \Phi) + \mat \Sigma_{1:}  \right)^{-1} {\mat V^{0,1:}(\vec \Phi)}^T 
\label{eq:v0}
\end{equation}

where $\vec R_{1:}$ is the column vector of dimension $\sum_{i=1}^p N_i$ concatenating $\vec y_1(\vec t_1) - \vec m_1(\vec t_1; \vec \theta_1), ..., \vec y_p(\vec t_p)- \vec m_p(\vec t_p; \vec \theta_p)$.  To simplify the analysis one can either:

\begin{enumerate}
    \item Instantiate the values of $\vec \Phi$ as maximising the likelihood describing constant stars radial velocity datasets $\vec y_1(\vec t_1),..., \vec y_p(\vec t_p)$ and consider $\vec \Phi$ as a constant in Equations \eqref{eq:mu0} and \eqref{eq:v0}. In this case, we compute a new master constant based on the GP prediction.
\newline
    
    \item Compute the posterior distribution of $\vec \Phi$ conditioned on $\vec y_1(\vec t_1),..., \vec y_p(\vec t_p)$ and use it as a prior in the analysis of the RV time series of interest, $y_0$, with covariance $\mat V_{0}'$ and mean $\vec m_0'$. This approach gives more conservative uncertainties.
\end{enumerate}

This second approach might also be directly implemented in the periodogram analyses of the $\ell_1$ \citep{2017MNRAS.464.1220H} to search for periodic signals in the data. We model the red noise in the $\ell_1$ as the covariance instantiated with the kernel whose parameters were adjusted on constant stars.

\section{Application to SOPHIE constant stars}
\label{section4}

To correct for NZP variations, the classical master constant was initially developed by \citet{2015A&A...581A..38C} based on four constant stars assumed to have no planets: HD 185144 \citep{2010Sci...330..653H}, HD 9407, HD 221354, and HD 89269A \citep{2011ApJ...730...10H}, along with 51 other targets that are monitored by SOPHIE. Recently, \citet{Heidari} updated the master constant using a larger sample of constant stars, including the initial four and all the SOPHIE Sub-programme 1 (SP1) targets \citep{2009A&A...505..853B} with low RV jitter and no detected periodic signal in their periodogram. We adopt the same sample of constant stars to train our GP-based NZP correction.

\subsection{Kernel selection}
\label{section_kernel}

\begin{figure}
     \centering
     \begin{subfigure}[b]{0.45\textwidth}
         \includegraphics[width=\textwidth]{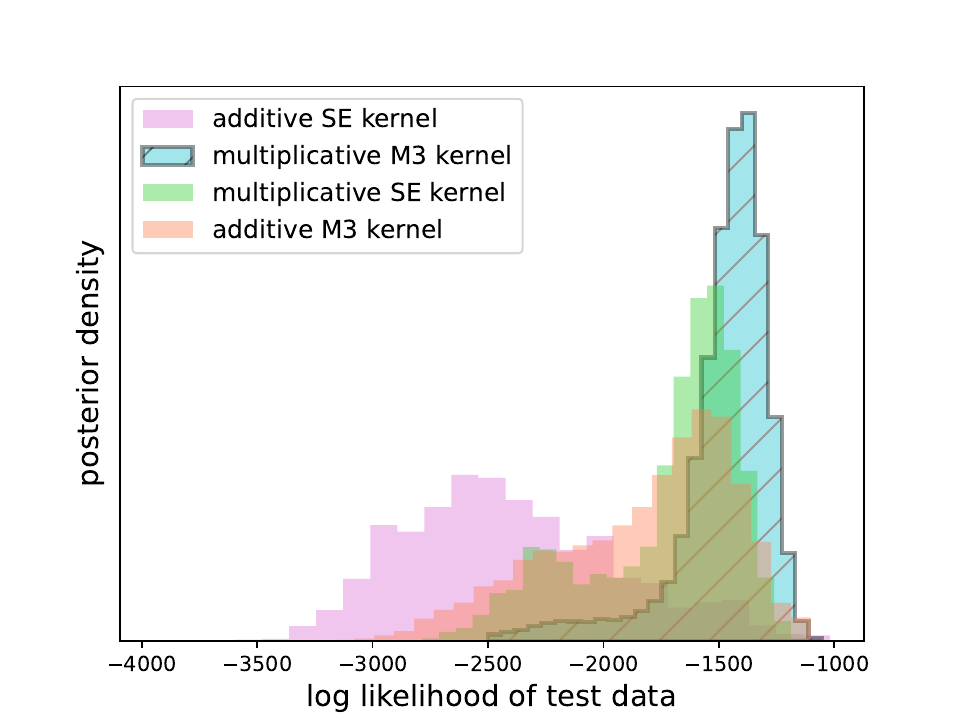}       
     \end{subfigure}
     \begin{subfigure}[b]{0.45\textwidth} 
        \includegraphics[width=\textwidth]{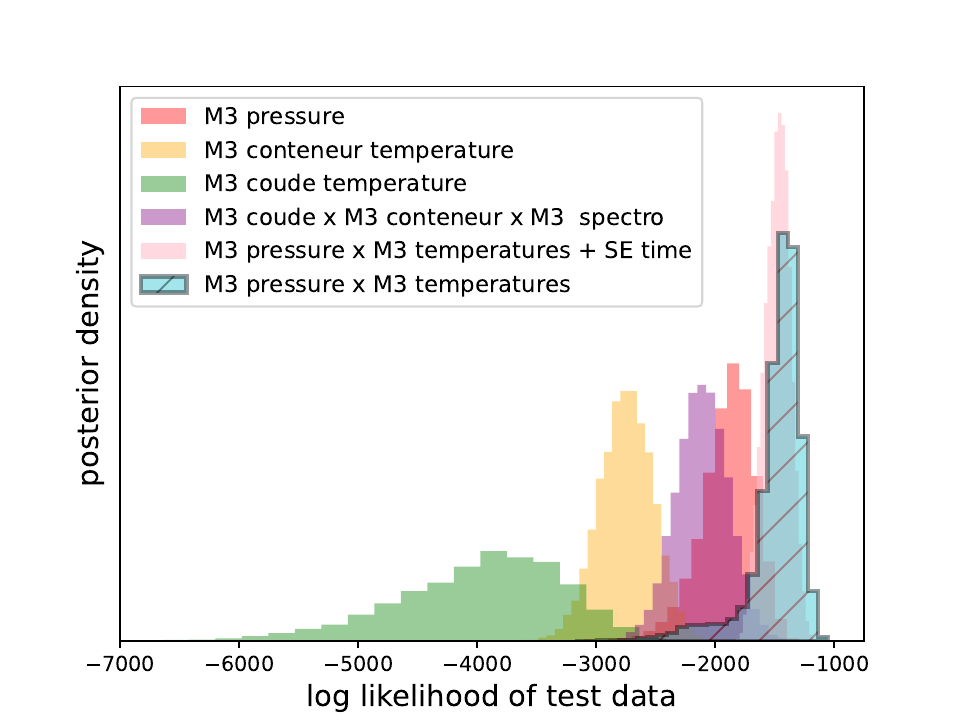}   
     \end{subfigure}
%          \begin{subfigure}[b]{0.4\textwidth}
%         \includegraphics[width=\textwidth]{pdfresizer.com-pdf-crop-1.pdf}       
%     \end{subfigure}
\caption{Results of the cross-validation method to select the best kernel. \textit{Upper panel:} We compared four kernels with multiplicative and additive kernels with both Matérn 3/2 (M3) or square-exponential (SE) kernels with the temperature of the air in the container, and the spectrograph, temperature of the coudé tube and pressure. The model that obtains the highest log-likelihood is the preferred one: multiplicative M3 kernel (MM3) in hashed blue. 
\textit{Lower panel}: Comparison of M3 or MM3 kernel with difference number of housekeeping variables. We compared models with only one temperature (yellow and green), only pressure (red), the three temperatures only (purple), the multiplicative M3 kernel with the 3 temperatures and pressure (blue) and the multiplicative M3 kernel with the addition of a SE kernel on time (pink).}
\label{histo_likelihood}
\end{figure}

   \begin{figure*}[h!]
   \centering
   %\sidecaption
   \includegraphics[width=\textwidth]{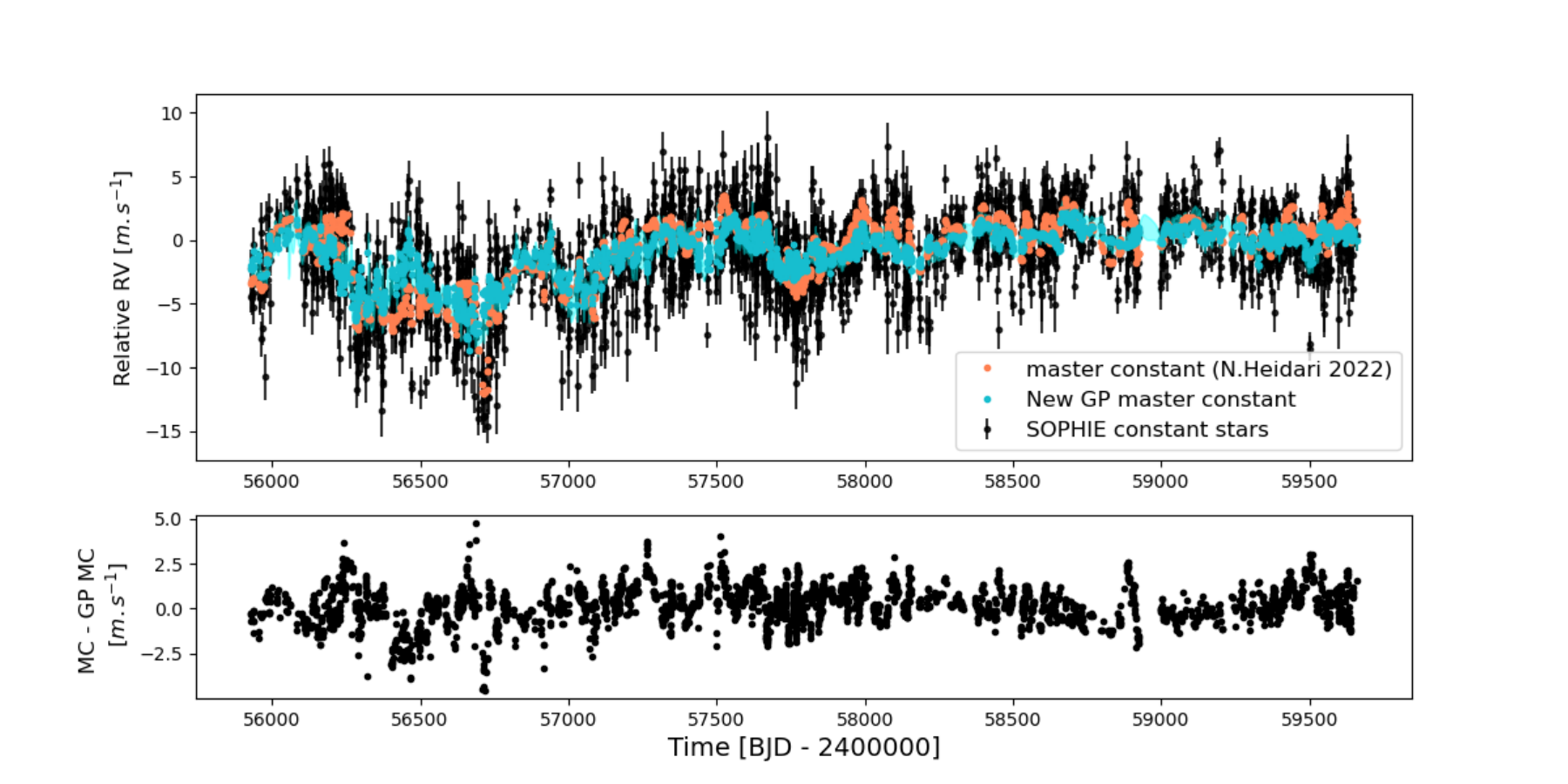}
      \caption{ Result of the fit of the NZP with GPs. \textit{Upper panel}: SOPHIE radial velocities for the constant stars are represented in black dots with errorbars. The master constant time series developed by \citet{Heidari} is represented as orange dots. The new GP master constant developed in this work and trained on the constant stars is represented with blue dots, and the blue shaded area represents the corresponding uncertainties. 
     \textit{Lower panel}: Difference between the master constant method (MC) and the GP master constant method (GP MC).
              }
         \label{fig_best_fit_params}
   \end{figure*}

The first step is to select which housekeeping variable will be used for the NZP correction. As explained in section \ref{section2}, some housekeeping variables show the same pattern and are likely different proxies of the same effect. Thus, we initially tested the air temperature in both the container and the spectrograph, the Coudé tube, and the cryostat. Since changes in pressure inside the spectrograph might cause a RV variation at the level of several hundred \ms\, \citep{2010EAS....41...27E}, we also tested the internal pressure in our initial list of housekeeping variables.

We fitted the sample of constant stars from \citet{Heidari} using the MCMC method described in section \ref{section3}. We used the five aforementioned housekeeping variables and both the additive (eq. \ref{addSE}) and multiplicative (eq. \ref{multSE}) SE kernels. Uniform priors were used for all hyperparameters. In both cases, the MCMC could constrain the hyperparameters for all housekeeping variables except for the cryostat. For the hyperparameter of the latter, the posterior distribution was similar to the one of the prior. Since its hyperparameter is unconstrained, we concluded that the cryostat has no significant impact on the NZP variations. Therefore, we excluded it from the following analysis while keeping the other four housekeeping variables: the internal pressure, the temperatures of the air in the container and spectrographs, and the Coudé tube.

The second step was to select the best kernel from the list presented in section \ref{subsection_kernels}, using the cross-validation method \citep[see chapter 5 of][]{Rasmussen}. We divided the RV data of the constant star sample into two sets: a training set with 67$\%$ of the data, selected randomly, and a test set with the remaining 33$\%$. The estimation of the hyperparameters of the four kernels using the MCMC algorithm is done on the training set. We then computed the likelihood of the test set at each step of the MCMC.
The upper panel of the figure \ref{histo_likelihood} displays the posterior distributions of the log-likelihood of the test set for each kernel. The multiplicative kernels (eqs. \ref{multSE} and \ref{multM3}) give a higher log-likelihood and are therefore preferred over the additives ones, especially for the SE kernel. This corroborates our hypothesis that all the housekeeping variables depend on each other. In this analysis, the multiplicative M3 kernel exhibits the highest log-likelihood, and thus, we adopt it. The Matérn class kernel offers a degree of variability of the process in between exponential kernels (non derivable realisations) and SE kernels (smooth realisations) \citep{Rasmussen}. 

The third step involved evaluating whether considering multiple housekeeping variables improved the analysis. We performed the same cross-validation as for the choice of the kernel but with different housekeeping variables. The lower panel of figure \ref{histo_likelihood} presents the results. The full model, selected with a multiplicative M3 kernel, is shown in blue. We also tested a model with only the temperature of the Coudé tube (green), only the temperature of the air in the container (yellow), and the three temperatures combined in a multiplicative M3 kernel (purple). Considering only one temperature yields a significantly lower log-likelihood than considering all three temperatures. The pressure alone (red histogram) performs better than using only temperature, confirming that pressure has a greater impact on the NZP variations. This analysis demonstrates that the highest log-likelihood is achieved with the full model. We also attempted to add a time-based SE kernel to the multiplicative M3 kernel based on the housekeeping variables, resulting in an equivalent log-likelihood.

As discussed in section \ref{section2}, the amplitude of the temperature and pressure variations decreases significantly near BJD=2457700 due to an upgrade of the thermal regulation in the spectrograph \citep{2018A&A...618A.103H}. This upgrade improved the instrumental stability, hence decreasing the NZP variations. Consequently, the covariance matrix between the RV time series and the housekeeping data should also be affected. Thus, we decided to split the analysis into two parts to consider this upgrade. The first fit with a MCMC of the constant star timeserie was therefore done only with a M3 multiplicative kernel on the three temperatures and pressure to avoid the over-flexibility of a time-based SE kernel.

The posterior distributions for each parameter are reported in the corner plots in figure \ref{corner_old} and \ref{corner_new} as blue histograms before and after the upgrade (respectively). The median and 68.3 \% credible intervals are specified for each parameter. We can see that the container's temperature variations strongly influence the RVs before the upgrade on the spectrograph but no longer after the upgrade. This is consistent with the upgrade, which focuses on the thermal regulation of the air inside the spectrograph (as seen in Fig. \ref{housekeeping_variables}) to better isolate the instrument from the air inside the container. 

 The final result of the MCMC for this GP noise model (eq. \ref{multM3}) is shown in figure \ref{fig_best_fit_params} where its prediction is compared with the master constant developed by \citet{Heidari}. The RMS of the residuals for the classical master constant correction is about 2.27~\ms, slightly larger than the RMS of the residual from our GP prediction with a value of 2.15~\ms. We managed to reproduce the main variations of the constant stars using only the pressure and three temperature variations. The difference between the classical master constant and the work presented here is shown in the lower panel of figure \ref{fig_best_fit_params}. At BJD~$\approx$~2456710, all the constant stars show a significant RV drop that is not reproduced by the GPs. This drop is contemporaneous with a change in the wavelength calibration of the spectrograph \citep{2018A&A...618A.103H} and thus is independent of the temperature or pressure inside the instrument.
 
\begin{figure}
     \centering
   \includegraphics[width=\columnwidth]{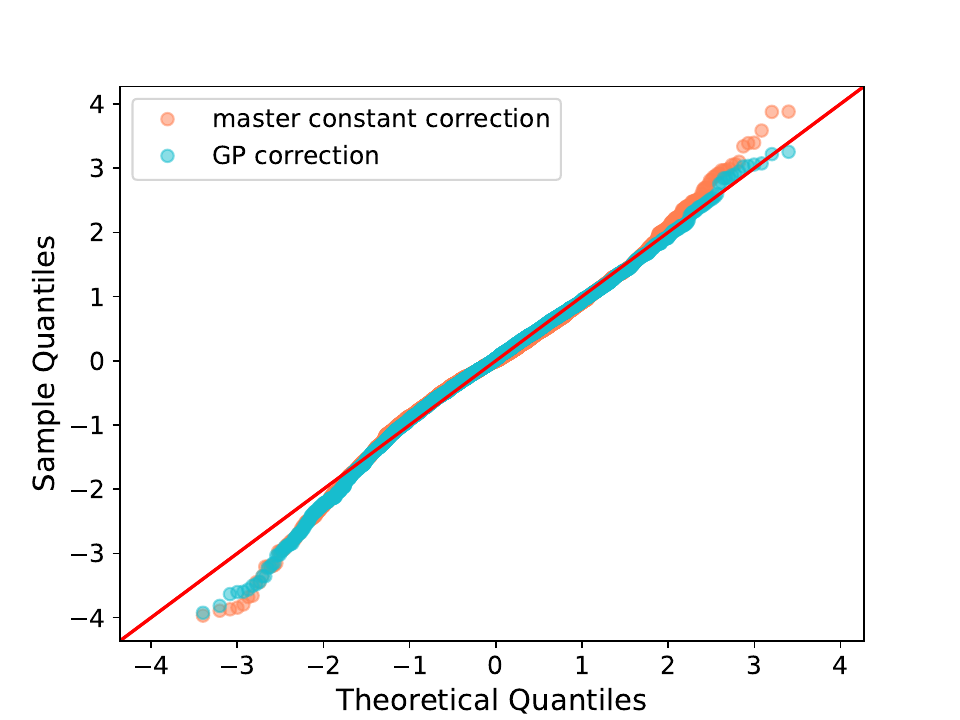}

\caption{Quantile - Quantile analysis comparing the quantile of the residuals for the master constant correction (orange dots) and the GP master constant correction (blue dots) with the corresponding theoretical quantile of a normal distribution. If the cumulative distribution of the residuals follows a normal distribution, the residuals should be aligned along a 45 \degr line (red line).}
\label{residuals}
\end{figure}

  Figure \ref{residuals} shows the quantile-quantile analysis of the residuals of both the classical master constant correction and the GP master constant from this work. The cumulative distribution function of the residuals is compared to the theoretical quantiles from a normal distribution. The analysis demonstrates that the GP master constant yields more Gaussian residuals, especially in the tail of the distribution, which supports the use of this method.

\begin{figure}
     \centering
   \includegraphics[width=\columnwidth]{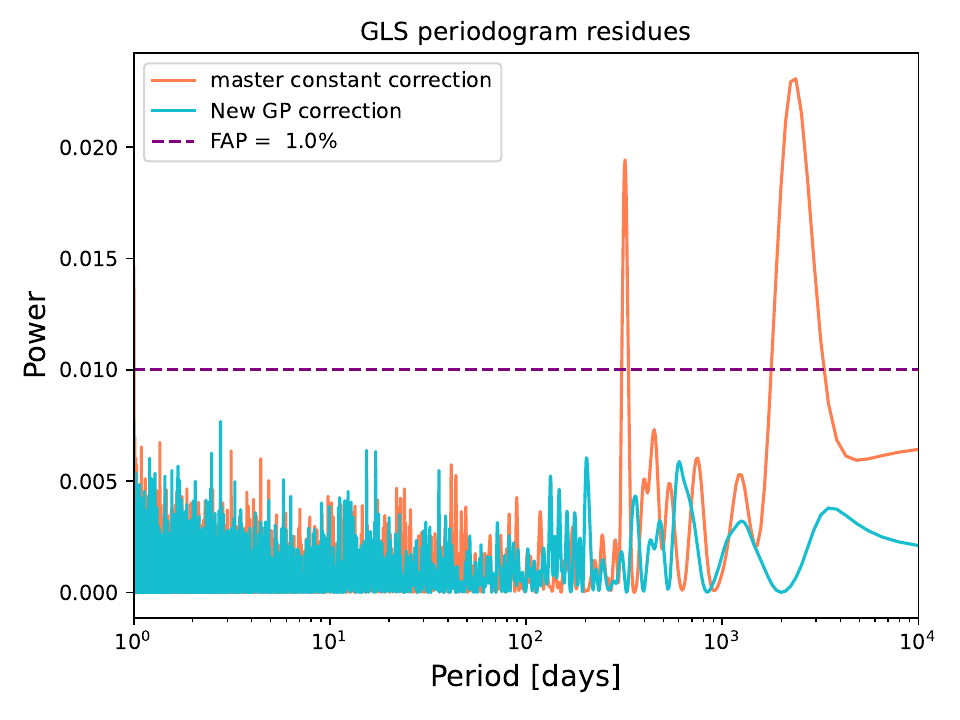}

\caption{Generalized Lomb-Scargle (GLS) periodogram \citep{2009A&A...496..577Z} of the residuals. The residuals of the NZP correction with the classical method are in orange, and the correction with the GP done in this work is in blue. The FAP at 1 $\%$ is represented as a dotted purple line.}
\label{periodogram_residuals}
\end{figure}

 Figure \ref{periodogram_residuals} compares the periodograms of the residues from the new GP master constant and the master constant correction \citep{Heidari}. No signal at a specific period is identified in the residuals from the method developed in this work. A peak at $\approx$ 360 days and one around 3000 days are significant in the classical method correction. As discussed in \citet{Heidari_phD}, these signals are likely due to instrumental variations and a potential long-period signal in one of the constant stars. Even if still present in the master constant correction, they have already been improved a lot since the last master constant developed in \citet{2015A&A...581A..38C}. The absence of these signals in the GP methods highlights that the latter should better correct the annual temperature and pressure variations. As no peak is identified in the periodogram of the residuals, we decided not to include other parameters in the MCMC.

\begin{figure}
     \centering
     \begin{subfigure}[b]{0.45\textwidth}
         \includegraphics[width=\textwidth]{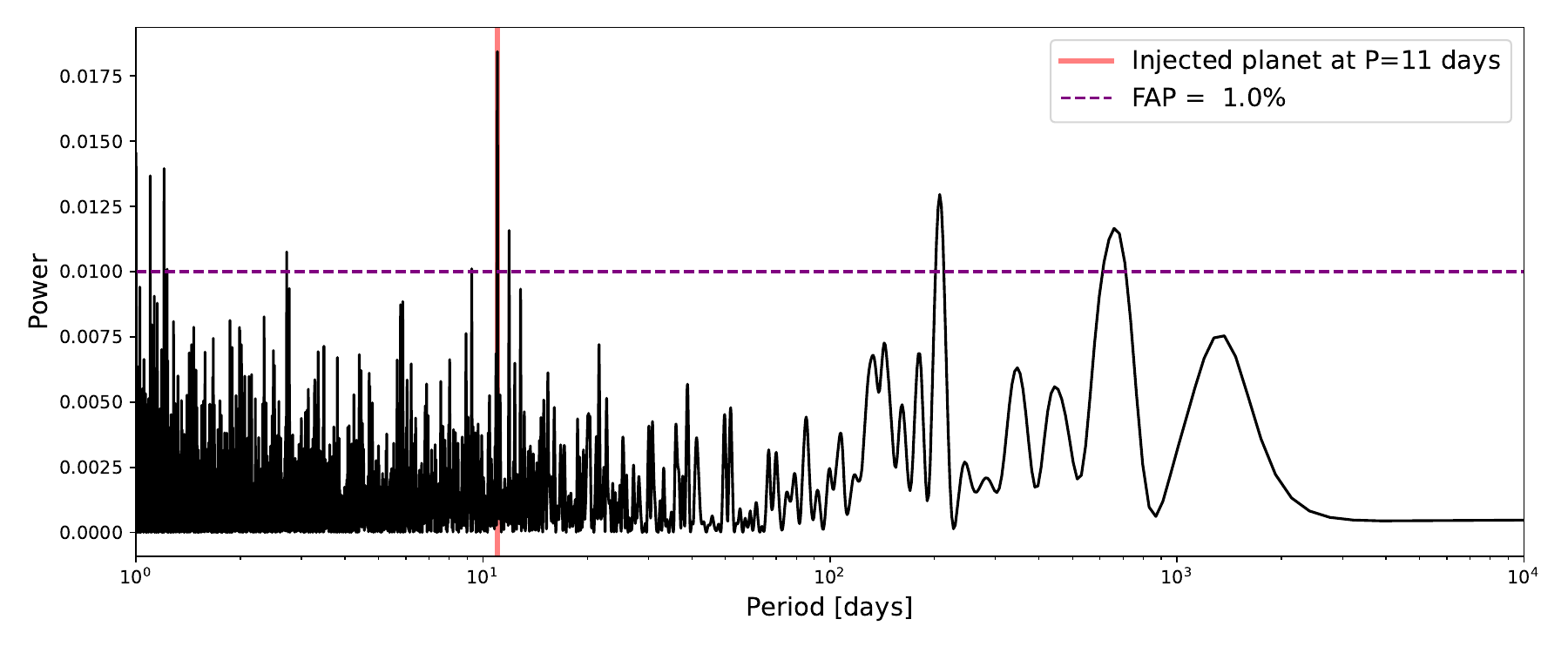}       
     \end{subfigure}
     \begin{subfigure}[b]{0.45\textwidth}
 
         \includegraphics[width=\textwidth]{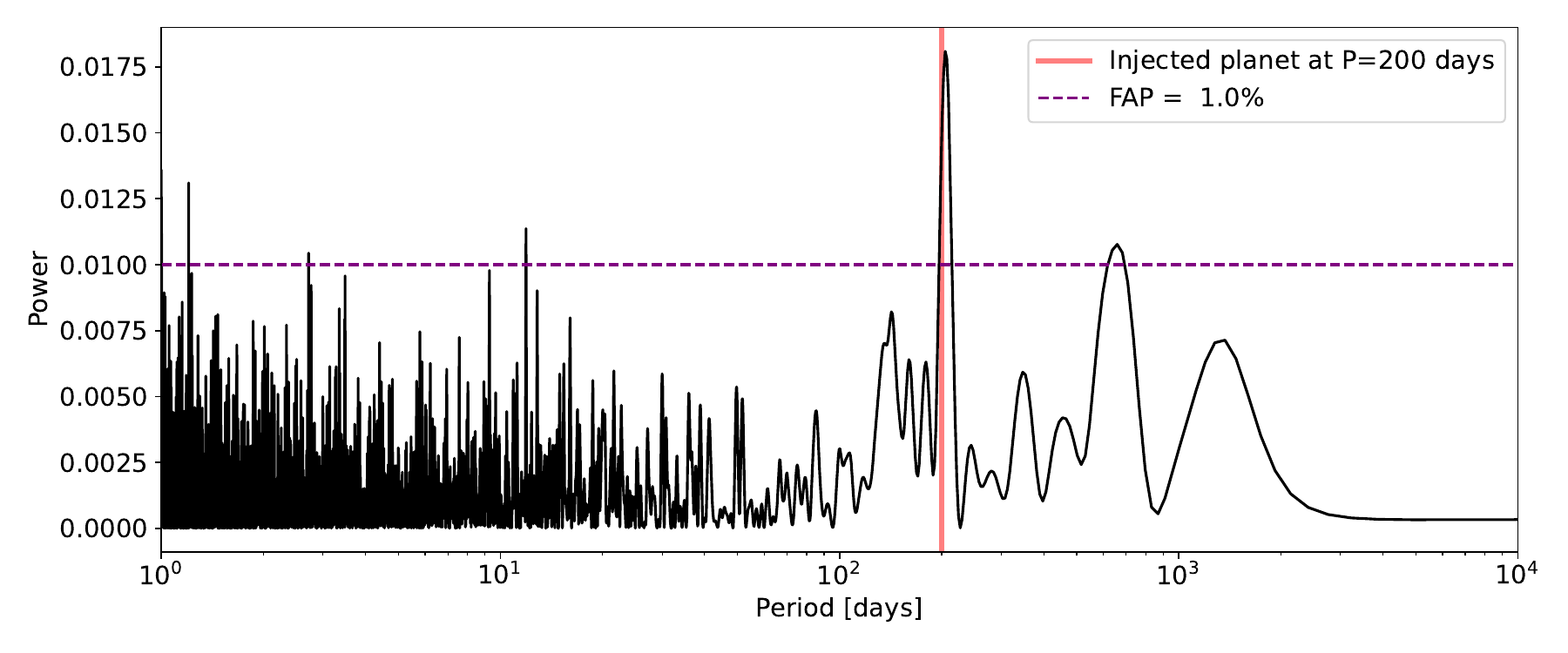}   
     \end{subfigure}
%          \begin{subfigure}[b]{0.4\textwidth}
%         \includegraphics[width=\textwidth]{pdfresizer.com-pdf-crop-1.pdf}       
%     \end{subfigure}
\caption{GLS periodograms of the residuals after the GP correction of the NZPs. Two planets have been injected in one of the constant stars of the total constant stars timeserie with an orbital period of 11 days (upper panel) and 200 days (lower panel). The FAP at 1 $\%$ is represented as a dotted purple line. The period of the injected planet is highlighted with a red vertical line.}
\label{fig_planets_in_cste_stars}
\end{figure}

As the Gaussian Processes are based on housekeeping variables, they should not model the Keplerian variation of a planet if it is present in the constant star. Even if constant stars show no significant variations over several years, the presence of a planet with a very small amplitude cannot be ruled out. Therefore, ensuring that the Gaussian Process does not consider such a planet is important.
To investigate this, we chose the constant star HD185144 and injected a Keplerian signal with an amplitude of 1 \ms and an orbital period of 11 days or 200 days. We then performed the same Gaussian Process analysis on the time series of all constant stars, including the one with the injected planet. Figure \ref{fig_planets_in_cste_stars} illustrates the results with two periodograms of the residuals. In both cases, the signal of the injected planet is detected, demonstrating that the Gaussian Process does not model a signal unrelated to pressure and temperature variations. These signals are also not present in the periodogram of the GP master constant and are not injected during the correction of the NZP of a star.

\subsection{Test on simulated planets}
\label{section_test}

   \begin{figure*}
   \centering
   %\sidecaption
   \includegraphics[width=\textwidth]{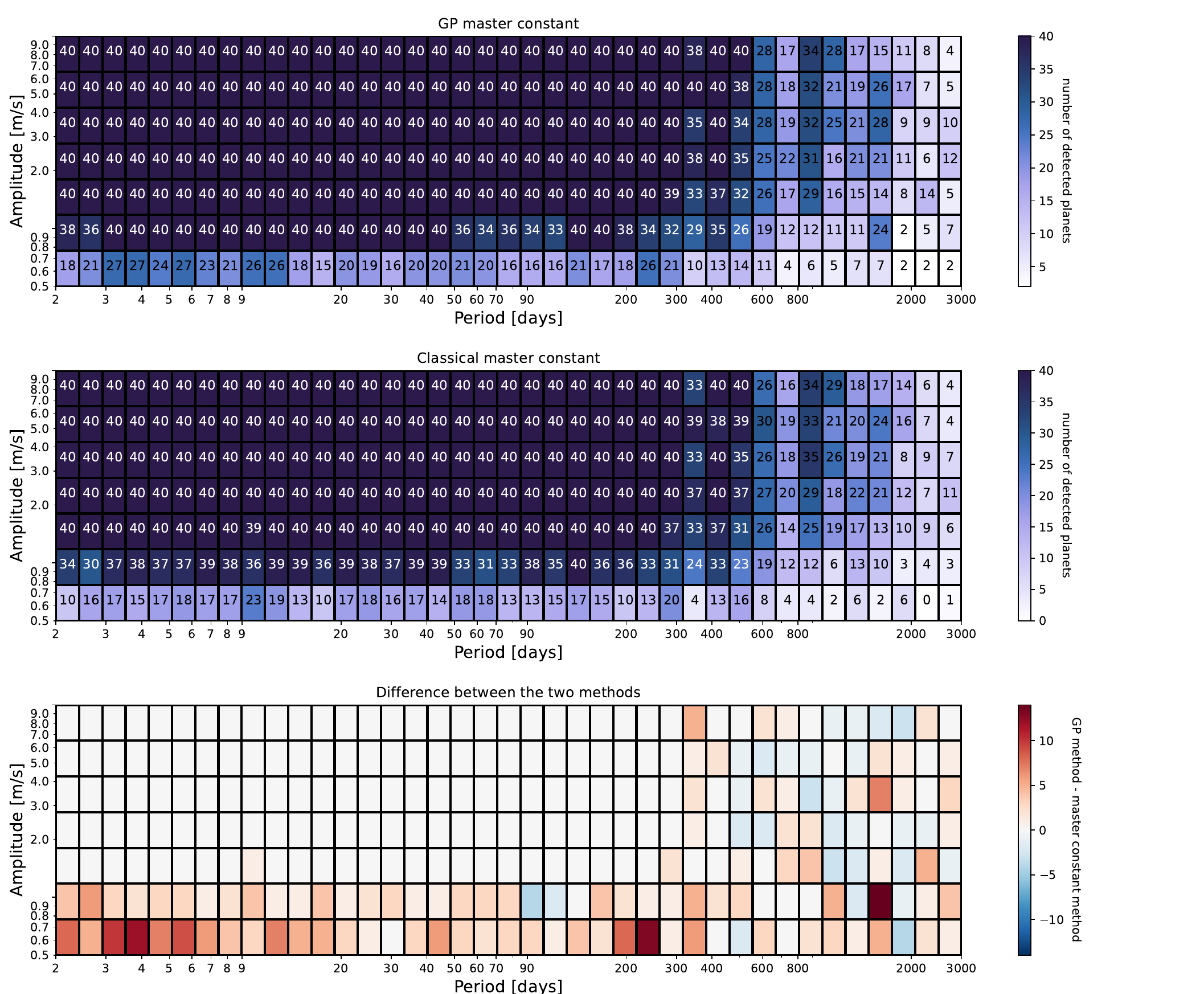}
      \caption{Number of planets detected as a function of the RV signal's period and semi-amplitude. The upper panel depicts the number of planets detected using the new GP master constant correction, and the second panel, the classical master constant correction. Grid cell colours indicate the amount of planets detected with the colour scale at the right. The darkest colour means more planets were detected. White cells contain no detected planets. The last panel shows the difference between the two methods regarding detected planets. Red colours mean more planets are detected with the GP method, and blue colours mean more planets are detected with the classical method. White cells mean that both methods detected the same number of planets. A planet is detected if $log_{10}(FAP) \le -1.3 $ and the orbital period is close to the injected one. }
         \label{fig_comparison_detection}
   \end{figure*}

To test the method for detecting and characterizing exoplanets, we simulated mock planets in the RV data of the constant star HD~89269A. We injected planets with periods ranging from 2 to 3000 days and semi-amplitude ranging from 0.5 to 10 \ms, both with a logarithmic scale. For each simulation, the eccentricity of the orbit is randomly chosen in a beta distribution following \citet{2013MNRAS.434L..51K} for multi-planetary systems and with a random argument of periastron between 0 and $2 \pi$. We evaluated the number of planets detected in a grid of cells in Figure \ref{fig_comparison_detection}. We generated 40 planets within each cell with random periods and semi-amplitudes within a given range. A planet is considered detected if its orbital period is valid with an error of $2 \%$ and its False-Alarm Probability \citep[FAPs, determined following the method described in][]{2020A&A...638A..95D} is $log_{10}(FAP) \le  -1.3 $. For each mock planet, we generated an l1-periodogram and considered only the most significant peak. We then counted the number of detected planets within each cell. We tested both the GP correction and the classical NZP correction for each simulated planet (first and second panels of Figure \ref{fig_comparison_detection}). This analysis aims to identify which method is more efficient given a region of parameters. The last panel of Fig. \ref{fig_comparison_detection} shows the difference between the two methods regarding the number of detected planets. A red cell is characterised by more detections by the GP method, and a blue cell has more detections by the classical methods.

Throughout the entire parameter space, the GP method detected more planets with $2 \%$ more planets using this method. Although this number is small because many planets are detected in both cases, the GP method is more efficient for small amplitudes. Within the range 0.5~-~1.2 \ms, the GP method detected 7.4 $\%$ more planets and even 10 $\%$ more in the range 0.5~-~0.7 \ms. For long periods, the classical method remains slightly better in some cases, primarily because GPs tend to lower the detection level of long-period signals. Therefore, the new GP instrumental correction could be used to detect small planetary signals. 

This study was done with a single-planet system only for simplicity, and we didn't use the GP noise model directly in the periodogram as it tends to lower the FAP of the signal.

When a planet is detected after the NZP correction, its orbital parameters should also be correct. We injected a 3-planets system in the constant star HD~89269A. The first planet has a semi-amplitude of 4 \ms, an orbital period of 35 days, and a circular orbit. The second planet has a semi-amplitude of 3 \ms, an orbital period of 90 days, and an eccentric orbit with $e=0.3$. The last planet has a semi-amplitude of 2.5 \ms\, and an orbital period of 264 days. All parameters are indicated in Table \ref{table_params_injected_planets}. All planets are clearly detected with a $log_{10}(FAP)~\le~-1.3 $. We then modelled these simulated data with an MCMC using the three methods for correcting the NZP variations. For all orbital parameters, we adopted large uniform priors, and we used a MCMC with 40 walkers with $3\cdot 10^{4}$ iterations and $4\cdot 10^{4}$ steps of burn-in. 

The results of the fits are provided in Table \ref{table_params_injected_planets}. For the three methods, the parameters are recovered within 1-$\sigma$. The classical master constant and the GP master constant show similar median and $68.3 \%$ credible intervals. The third method is different as it models the covariance matrix directly within the MCMC, although the hyperparameters are constrained within the posterior of the constant star training. As discussed in \citet{2015MNRAS.451..680E}, this approach is much faster than using large and uniform priors for the hyperparameters, as GPs are computationally expensive. The derived uncertainties are, therefore, lower limits of the true uncertainties but are more accurate than the ones obtained with the master constant method where the uncertainties are not propagated. As reported in table \ref{table_params_injected_planets}, the uncertainties for all the periods and amplitude parameters are between ~1.5 and 2 times larger with the GP noise model. For the eccentricities and argument of periastron, the uncertainties are equivalent. The GPs reflect our ignorance of instrumental systematics and produce more consistent uncertainties on the planetary parameters. 

\subsection{Gaussian Processes predictions}
\label{section_prediction}

Four constant stars are monitored with SOPHIE every possible night. This monitoring decreases the time allocated to the search for exoplanets. We have developed a new method that can predict the variations of the NZP using only a few housekeeping variables. As a result, decreasing the observations of constant stars may be possible.

To test this, we produced three different datasets. One with the complete time series of constant stars introduced in section \ref{section_kernel}, one with 20$\%$ of the data randomly removed from the complete time series, and a last one with 40$\%$ of the data randomly removed. We decrease the number of observations but keep the number of constant stars observed to have observations all over the years. We then applied exactly the same method described in section \ref{section_GP_model} to these new datasets. As before, we divided the constant stars dataset in two to account for the thermal regulation upgrade (at BJD $\sim$ 2457700). To compare the probability distribution of each hyperparameter, we produced two corner plots that show the posterior distributions of the hyperparameters for all three datasets. The two correlation diagrams are in Appendix \ref{corner_old} and \ref{corner_new}. The posterior distributions before and after the instrumental upgrade are consistent. The set with40$\%$ of the timestamps of all constant stars removed shows some divergences, particularly for the temperature of the container length-scale hyperparameter ($l_{container}$). This result is promising, as it indicates that even with fewer constant star observations, we can still produce a noise model that explains most of the NZP variations. Monitoring constant stars is still necessary to train the GP accurately, but it can be done at a lower cadence.

To test if the cadence decrease impacts planetary signal detection, we tested these two GP master constants with fewer data points with our mock planetary system injected into HD~89269A. In all three cases, the RMS of the residuals are at the level of 3.6~\ms. Concerning the detection level, we also used the $\ell_1$ and compute the FAPs of each signal. Even with $40\%$ less observations of the constant stars data, the three mock planets are fully recovered, and their FAPs are only marginally lower than those found in section \ref{section_test}.

\section{Application to planetary systems}
\label{section5}

In this section, we test our new method to correct NZP variations on a known planetary system: HD~158259 which hosts six exoplanets in a 3:2 mean-motion resonance \citep{2020A&A...636L...6H}. This detection was sensitive to the noise model, and we want to compare it with other instrumental systematical corrections. Five planets have been detected in RVs at 3.4, 5.2, 7.9, 12 days and another at 1.84 or 2.17 days, which are aliases of each other. Another candidate planet has been detected at 17.4 days. The internal planet at 2.17 days has been confirmed in transit with TESS observations. 
\citet{2020A&A...636L...6H} used the master constant developed at that time and tested new noise models. They selected the best noise model based on cross-validation. We compared the published FAPs with three other corrections: first, using the same noise model as in \citet{2020A&A...636L...6H} but with the master constant by \citet{Heidari}, then using the new GP master constant developed in this work, and finally, using the raw data without any master constant correction, and applying our noise model directly in the $\ell_1$. 

\begin{figure}
     \centering
   \includegraphics[width=\columnwidth]{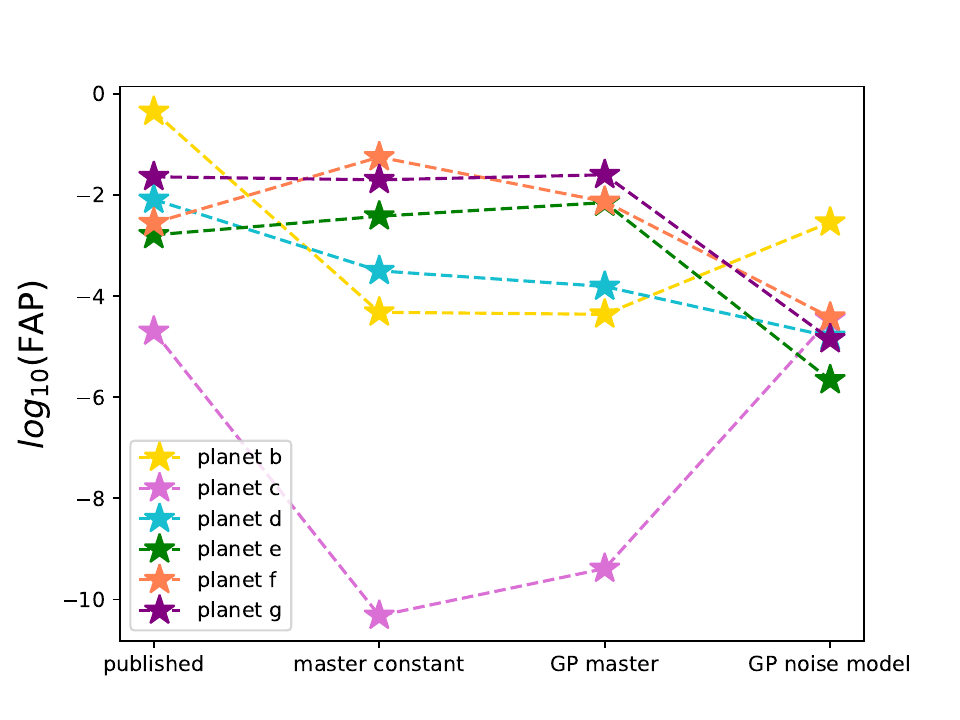}

\caption{$\log_{10}(FAP)$ (False Alarm Probability) for each instrumental variation correction of HD~158259 RVs. In the y-axis, we show the four different corrections: The published one, the master constant developed by \citet{Heidari}, the new GP master constant, and raw data with the GP noise model. Coloured stars represent the FAPs, and each planet has a colour: yellow stars for planet b, pink for planet c, blue for planet d, green for planet e, orange for planet f, and purple for planet g. }
\label{HD158259_FAP}
\end{figure}

The results for each planet are shown in figure \ref{HD158259_FAP}. The $\log_{10}$(FAP) are reported for each method, and we see that the results are quite different according to the planets. Planet b is the only one confirmed by transit, and we see a significant improvement in the detection for the three methods compared to the published one. The master constant developed by \citet{Heidari} and the GP master constant show similar results, but the noise model show a higher FAP. The situation is similar for planet c at 3.4 days. However, for the other planets (d, e, f, and g), the GP noise model significantly improved the detection, with planet g now significantly detected. The application of the GP noise model on this system shows that we recover all the planets, with equal or better FAP, without using the red noise model used in \citep{2020A&A...636L...6H}. The period of planet b found with the GP correction is equal to that detected with TESS and not to its alias of 1.84 d as in \citep{2020A&A...636L...6H}.

\begin{figure*}
     \centering
   \includegraphics[width=\textwidth]{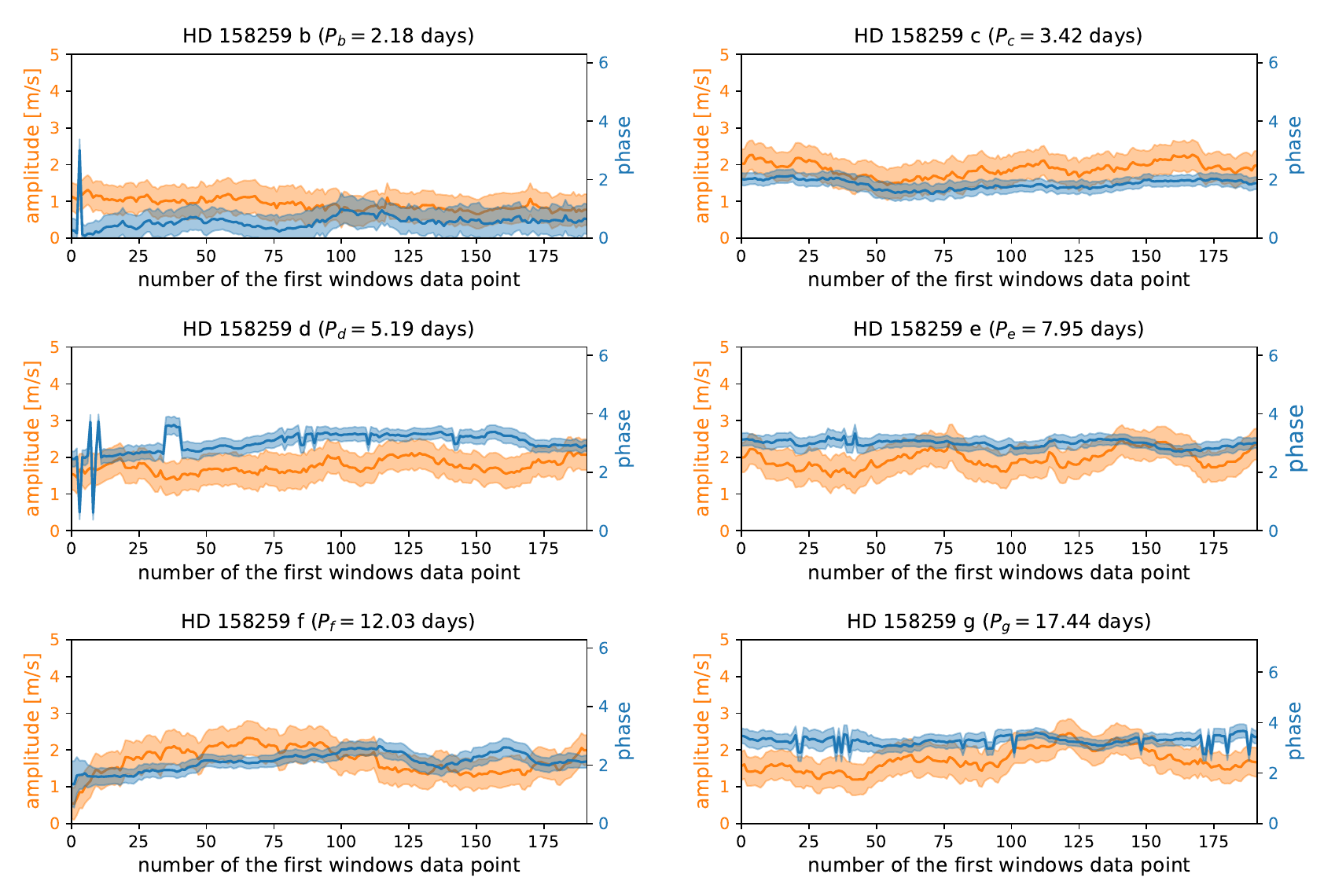}

\caption{Evaluation of the stability of the signal over time for the six planets in HD~158259. The orbital period of each planet was fixed to the one found with the MCMC and written above each panel. The semi-amplitude (orange) and phase (blue) are estimated in a window of 60 RV data points that we shifted points by points. The shaded area shows the uncertainties.}
\label{amplitude_phase}
\end{figure*}

We performed a MCMC to model the 6 planets of this system. In table \ref{table_params_HD158259}, we compared the posteriors obtained with a GP noise model with the posteriors from \citep{2020A&A...636L...6H}. We used the same priors and stellar mass as in \citep{2020A&A...636L...6H}. With a noise model only based on pressure and temperatures and the addition of a jitter, we found compatible orbital parameters for all the planets. In this case, the noise model is based on instrumental conditions. To confirm the signals of the planets, we evaluated if the six planetary signals are coherent over time. We considered a window of 60 RV data points that we shifted points by points until the end of the timeserie. We evaluated the amplitude of the signal and the phase of each planet, considering a circular orbit, in each window. The result is shown in figure \ref{amplitude_phase} and follows the method in \citet{2022A&A...658A.177H}. For the six planets, the semi-amplitude and phase are constant over the observations, reinforcing that these signals are constant in time and have a planetary origin. With this study, the hypothesis that HD~158259g is a planet is strengthened and will be fully confirmed with more observations in a follow-up paper.

\section{Discussion and conclusion}
\label{section6}

High-precision spectrographs are crucial for detecting and characterising exoplanets. This paper introduces an improved method to understand and correct the NZPs variations on the SOPHIE spectrograph. While the SOPHIE spectrograph is stabilised in pressure and temperature, it is not placed in a vacuum chamber. By studying the NZP pattern using constant stars monitored with SOPHIE, we aim to overcome instrumental instabilities. Additionally, we utilise ancillary information from the instrument, namely pressure and temperature variations, which are measured every 6 minutes. Previously, these parameters have not been used to correct NZP variations. GPs are employed to analyse the impact of pressure and temperature variations on the RV time series. GPs allow for modelling correlated noise and accurately propagating uncertainties in planetary parameters. The product of multivariate Matérn-3/2 kernels is suitable to model these NZP variations. This study allows us to understand the influence of environmental conditions on SOPHIE RV measurements. The flexibility of the GPs allows us to handle complex correlations where we don't have \textit{apriori} knowledge of the impact of external variations on the data.

We used the same set of constant stars as the classical master constant correction. These stars have been monitored for over a decade since 2012. To correct for the NZP variations, we propose two approaches based on the housekeeping data: (1) the classical master constant is replaced by a GP master constant from the prediction of the best kernel fitted on the constant stars, and (2) a noise model that can be self consistently implemented within MCMC or the $\ell_1$ with the hyperparameters trained on the constant stars. The GP master constant with associated uncertainties is provided in the Appendices (Table \ref{table_GP_master}). One main limitation of this method is that GPs are computationally expensive.

Using GPs to correct NZP leads to detecting more planets with small amplitudes. We showed that up to $10 \%$ more planets could be detected with an amplitude below 1 \ms. Therefore, the GP method has the potential to unveil smaller mass planets across a wide range of orbital periods. For larger amplitude, both the GP master constant and the classical master constant developed by \citet{Heidari} have similar efficiency in correcting for NZP variations.

As already discussed in section \ref{section5} and in \citep{2020A&A...636L...6H}, comparing different noise models is important to assess the validity of planet detection. We applied the method on a complex planetary system HD~158259 with small amplitude signals. The new correction of the NZP variations helps us to confirm the planets b to f of the system and reinforce the possibility of the signal at 17.44 days being a planet.
We discuss in section \ref{section_prediction} that we could reduce the observation's cadence of the constant stars without affecting the results. However, completely discontinuing their monitoring and relying solely on housekeeping data might be too risky. Indeed, temperature and pressure variations cannot explain all the NZP variations, as we already mentioned with the update of the wavelength calibration (end of section \ref{section_kernel}). 

RV analysis is a complex process, and various external factors and steps in the reduction pipeline can introduce instrumental systematics. Factors such as seeing variations, observational conditions, Barycentic Earth Radial Velocity, opto-mechanical variations of the instrument, and other effects must be considered to fully understand instrumental variability. While additional ancillary information could be incorporated into our GP-based method, computational cost and parameter degeneracy remain primary limitations. Nonetheless, using pressure and temperature variations alone, we successfully capture the main variations of NZP. This method will be applied to other planetary systems discovered with SOPHIE.

\newpage

\begin{acknowledgements}
The project leading to this publication has received funding from the Excellence Initiative of Aix-Marseille University - A*Midex, a French “Investissements d’Avenir programme” AMX-19-IET-013. This work was supported by the "Programme National de Planétologie" (PNP) of CNRS/INSU.
\end{acknowledgements}

%%%%%%%%%%%%%%%%%%%%%%%%%%%%%%%%%%%%%%%%
% BIBLIOGRAPHY
%%%%%%%%%%%%%%%%%%%%%%%%%%%%%%%%%%%%%%%%

\bibliographystyle{aa} 
\bibliography{bibtex}

\appendix
\onecolumn
\section{Supplement figures and tables}
%\begin{appendix}

\begin{figure*}[h]
     \centering
   \includegraphics[width=\textwidth]{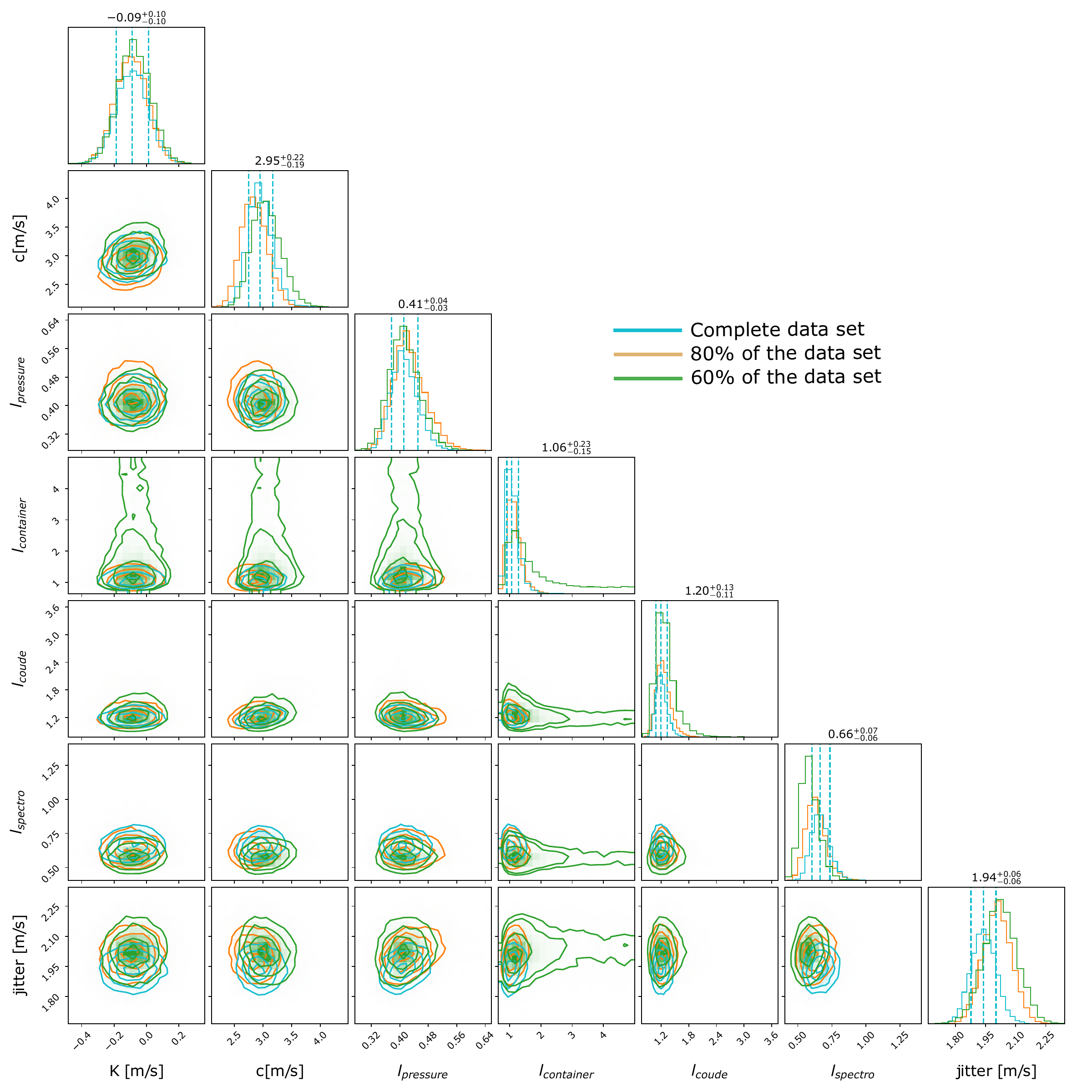}

\caption{Correlation diagram for the posterior density functions of all parameters for the GP analysis of the constant stars with BJD<2457700 days. The blue histogram corresponds to the complete data set with all constant stars available. The orange histograms are the results for 80$\%$ of this data set, and the green histogram for $60\%$. The blue dotted lines show the median values for the complete data set and limit the  $68.3\%$ highest density intervals. The corresponding values are written above each parameter. }
\label{corner_old}
\end{figure*}

\begin{figure*}
     \centering
   \includegraphics[width=\textwidth]{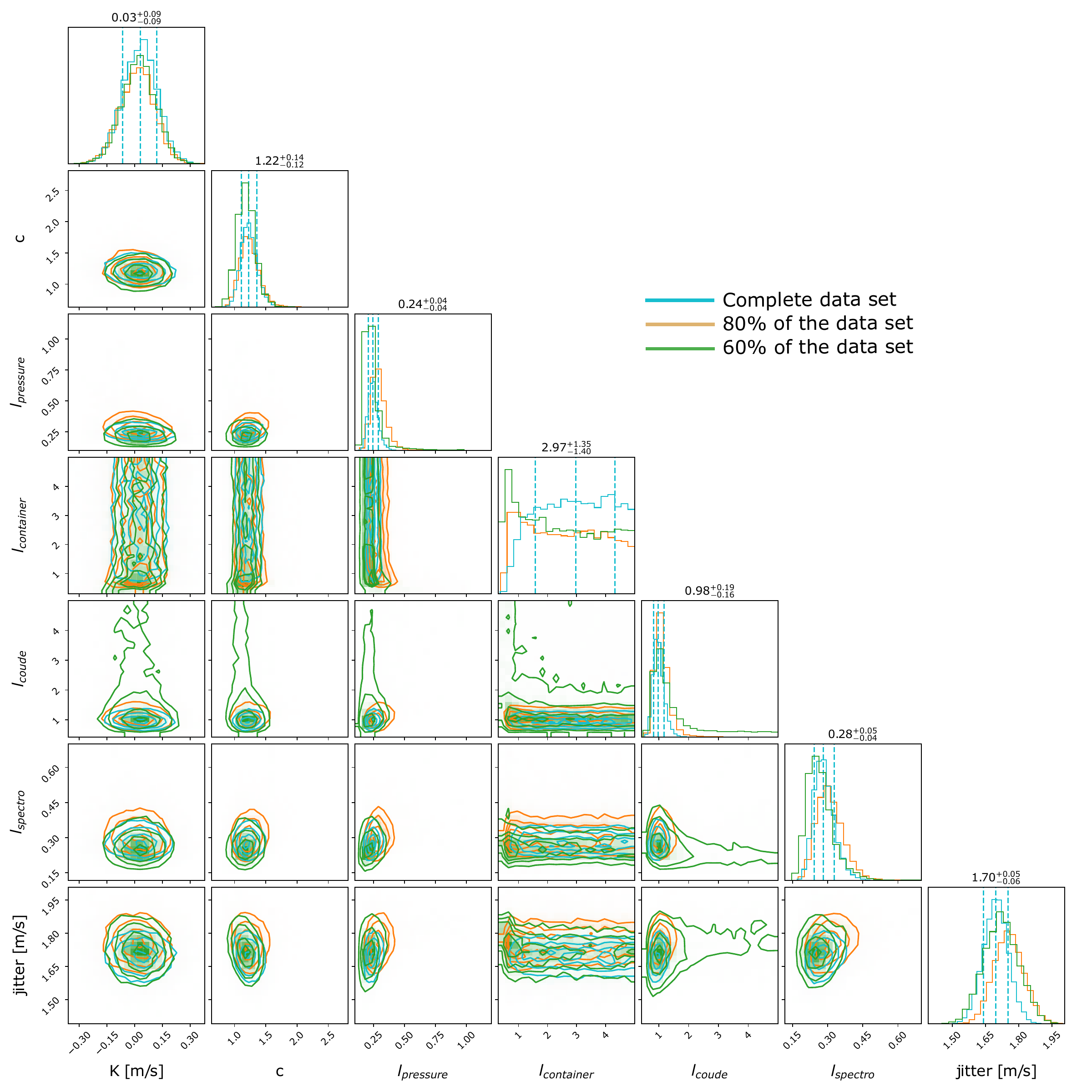}

\caption{Correlation diagram for the posterior density functions of all parameters for the GP analysis of the constant stars with BJD>2457700 days. The blue histogram corresponds to the complete data set with all constant stars available. The orange histograms are the results for 80$\%$ of this data set, and the green histogram for $60\%$. The blue dotted lines show the median values for the complete data set and limit the  $68.3\%$ highest density intervals. The corresponding values are written above each parameter.}
\label{corner_new}
\end{figure*}

\begin{table*}
\centering
\caption{List of implemented and fitted parameters for planets injected in HD~89269A.}
\begin{tabular}{lcccc}
\hline 
parameter & true value & master constant fit & GP master constant fit & GP noise model  \\ 
\hline 
%$T_{0;b} [BJD]$ && $57858.810_{-0.080}^{+0.092}$& $57858.936_{-0.078}^{+0.093}$  \\ 
%$T_{0;c} [BJD]$ &57921& $57921.19\pm0.94$ &$57921.05\pm0.66$ \\ 
%$T_{0;d} [BJD]$ &57942& $57943.9_{-4.6}^{+5.5}$ & $57943.4_{-3.9}^{+3.5}$ \\ 
$P_b$ [d] &35& $34.995 \pm 0.006 $ &$34.988\pm0.007$ & $34.986\pm0.009$ \\ 
$P_c$ [d] &90& $90.05_{-0.05}^{+0.05}$ &$90.03\pm0.05$ &$90.08\pm0.06$\\ 
$P_d$ [d] &264& $264.1\pm 0.5$ &$263.9\pm0.6$ &$262.6\pm1.1$ \\ 
$K_b$ [\ms] &4& $3.8\pm0.1$ &$3.7\pm0.1$ & $3.7_{-0.2}^{+0.2}$\\ 
$K_c$ [\ms] &3& $2.9\pm0.1$ &$3.1\pm0.1$ & $3.1\pm0.2$\\ 
$K_d$ [\ms] &2.5& $2.9 \pm 0.1$ &$2.8 \pm 0.1$ & $2.7 \pm 0.3$\\ 
$\sqrt{e}\cos{\omega}_b$ &0& $0.04_{-0.03}^{+0.02}$ &$0.04_{-0.02}^{+0.04}$ &$0.04_{-0.02}^{+0.04}$  \\ 
$\sqrt{e}\sin{\omega}_b$ &0& $0.02_{-0.01}^{+0.02}$ &$0.02_{-0.03}^{+0.04}$  &$0.02_{-0.01}^{+0.02}$\\ 
$\sqrt{e}\cos{\omega}_c$ &0.274& $0.11 \pm 0.03$ &$0.11\pm0.06$  &$0.10\pm0.04$ \\ 
$\sqrt{e}\sin{\omega}_c$ &0.474& $0.483 \pm 0.027$ &$0.485 \pm 0.021$ &$0.482_{-0.026}^{+0.023}$  \\ 
$\sqrt{e}\cos{\omega}_d$ &0& $0.07_{-0.03}^{+0.04}$ &$0.05_{-0.03}^{+0.04}$ &$0.05_{-0.05}^{+0.03}$ \\ 
$\sqrt{e}\sin{\omega}_d$ &0& $0.002\pm0.001$ &$-0.002_{-0.002}^{+0.004}$ & $0.002\pm0.002$  \\ 
\hline 
\end{tabular}
\label{table_params_injected_planets}
\end{table*}

\begin{table*}
\centering
\begin{threeparttable}
\caption{List of the GPs hyperparameters used in the analysis.}
\begin{tabular}{lccc}
\hline 
parameter & Prior  & \multicolumn{2}{c}{Posterior (median and 68.3\% C.I.)  } \\ & & BJD < 57700  & BJD > 57700\\
\hline 

$c$ [\ms]& $\mathcal{U}(0,5)$ & $2.946_{-0.194}^{+0.221}$ & $1.223_{-0.117}^{+0.137}$\\ 
\\
$l_{\mathrm{pressure}}$  & $\mathcal{U}(0,5)$ & $0.411_{-0.034}^{+0.040}$ & $0.243_{-0.036}^{+0.043}$ \\ 
\\
$l_{\mathrm{container}}$ & $\mathcal{U}(0,5)$  & $1.060_{-0.149}^{+0.226}$ & $2.974_{-1.405}^{+0.043}$ \\ 
\\
$l_{\mathrm{coude}}$ & $\mathcal{U}(0,5)$  & $1.197_{-0.108}^{+0.134}$ & $0.978_{-0.155}^{+0.192}$ \\ 
\\
$l_{\mathrm{spectro}}$ & $\mathcal{U}(0,5)$  & $0.665_{-0.062}^{+0.071}$ & $0.281_{-0.038}^{+0.047}$ \\ 

\hline 
\end{tabular}
\label{table_params_GP}

\begin{tablenotes}
      \small

        \item  Values for the length scale hyperparameters $l$ have arbitrary normalization \\ due to 
the standardization of their associated housekeeping variables.
      
    \end{tablenotes}
\end{threeparttable}
\end{table*}

\begin{table*}
\centering
\begin{threeparttable}
\caption{GP master constant values }    
%\begin{center}
\begin{tabular}{ccc}
\hline
Time {[BJD-2400000]} & RV [$m.s^{-1}]$          & $\sigma_{RV}$ [$m.s^{-1}$]    \\ \hline

55929.3700  & -2.26712 & 0.82183 \\
55929.6399  & -2.33950 & 0.76538 \\
55930.3600  & -2.65565 & 0.78356 \\
55930.4199  & -2.65309 & 0.77646 \\
55930.5999  & -2.52369 & 0.95596 \\
55934.6699  & -1.93319 & 0.10347 \\
55935.6699  & -1.86598 & 0.76388 \\
55936.6399  & -2.42267 & 0.79469 \\
55937.6699  & -1.25189 & 0.13099 \\
55938.2399  & -2.44164 & 0.10886 \\
55938.3000  & -2.43332 & 0.10761 \\
...         &...       &  ...      \\
...         &...       &  ...      \\
...         &...       &  ...      \\
59644.6200  & 0.70019  & 0.50288  \\
59645.2799  & 0.17656  & 0.63796  \\
59645.5299  & 0.56003  & 0.50445  \\
59646.4700  & 0.47732  & 0.48548  \\
59648.5599  & 0.49168  & 0.47227  \\
59658.4400  & -0.0027  & 0.69944  \\
\hline

\end{tabular}

\label{table_GP_master}
\begin{tablenotes}
      \small

        \item  Full table available 
      
    \end{tablenotes}
\end{threeparttable}
\end{table*}

\begin{table*}
\caption{Estimate and credible interval of the orbital parameters for the 6 planets of HD~158259 and a GP noise model compared with the published values in \citep{2020A&A...636L...6H}}
\begin{tabular}{lccc}
\hline 
parameter model &Posteriors from {\citet{2020A&A...636L...6H}}& Posterior (model with GP noise model)  & 99.73 \% confidence interval  \\ 
\hline 
\textit{Planet b} & & &\\
$P_b$ [d] & $2.178_{-0.0001}^{+0.00009}$ &  fixed at 2.178    & fixed \\ 
$K_b$ [m/s] & $1.05_{-0.19}^{+0.18}$&  $0.91_{-0.22}^{+0.25}$    & [0.50,1.59] \\ 
$\sqrt{e_b}\cos{\omega_b}$  & $0.08_{-0.21}^{+0.37}$&  $0.03_{-0.15}^{+0.17}$    & [-0.36,0.47] \\ 
$\sqrt{e_b}\sin{\omega_b}$ & $0.02_{-0.18}^{+0.24}$ &  $0.01_{-0.15}^{+0.16}$    & [-0.42,0.47] \\ 
$m_b \sin{i_b}$ [\Mearth] & $2.21_{-0.44}^{+0.40}$&  $1.95_{-0.47}^{+0.53}$   & [1.08,3.40] \\ 
\hline 
\textit{Planet c} & & &\\
$P_c$ [d] & $3.432_{-0.0002}^{+0.0003}$&  $3.4318_{-0.0003}^{+0.0004}$   & [3.4308,3.4327] \\ 
$K_c$ [m/s] & $2.26_{-0.20}^{+0.19}$ &  $2.04_{-0.26}^{+0.27}$    & [1.28,2.73] \\ 
$\sqrt{e_c}\cos{\omega_c}$  & $-0.01_{-0.13}^{+0.11}$&  $-0.05_{-0.14}^{+0.14}$    & [-0.444, 0.306] \\ 
$\sqrt{e_c}\sin{\omega_c}$  & $-0.0009_{-0.1291}^{+0.1191}$&  $-0.03_{-0.15}^{+0.15}$    & [-0.47,0.357] \\ 
$m_c \sin{i_c}$ [\Mearth] & $5.59 _{-0.58}^{+0.61}$&  $5.08 _{-0.64}^{+0.65}$   & [3.19,7.84] \\ 
\hline 
\textit{Planet d} & & &\\
$P_d$ [d] & $5.1981_{-0.0008}^{+0.0008}$&  $5.1986_{-0.0009}^{+0.0009}$   & [5.1957,5.2011] \\ 
$K_d$ [m/s]  & $2.26_{-0.20}^{+0.19}$&  $1.95_{-0.26}^{+0.27}$   & [1.17,2.75] \\ 
$\sqrt{e_d}\cos{\omega_d}$  & $0.12_{-0.16}^{+0.17}$&  $0.06_{-0.14}^{+0.15}$   & [-0.29,0.48] \\ 
$\sqrt{e_d}\sin{\omega_d}$   & $-0.03_{-0.15}^{+0.15}$&   $-0.007_{-0.135}^{+0.145}$  & [-0.384,0.378] \\ 
$m_d \sin{i_d}$ [\Mearth] & $5.39 _{-0.69}^{+0.76}$&  $5.54_{-0.75}^{+0.79}$  & [3.34,7.85] \\ 
\hline 
\textit{Planet e} & & &\\
$P_e$ [d]& $7.951_{-0.002}^{+0.002}$ &  $7.949_{-0.002}^{+0.002}$   & [7.943,7.954] \\ 
$K_e$ [m/s] & $1.86_{-0.28}^{+0.27}$&  $1.92_{-0.27}^{+0.29}$    & [1.13,2.71] \\ 
$\sqrt{e_e}\cos{\omega_e}$   & $-0.07_{-0.14}^{+0.15}$&  $-0.03_{-0.15}^{+0.15}$  & [-0.44,0.38] \\ 
$\sqrt{e_e}\sin{\omega_e}$ & $0.03_{-0.14}^{+0.14}$ &   $0.05_{-0.13}^{+0.18}$   & [-0.32,0.46] \\ 
$m_e \sin{i_e}$ [\Mearth] & $6.06_{-1.01}^{+0.96}$&  $6.32_{-0.88}^{+0.94}$   & [3.69,8.91] \\ 
\hline 
\textit{Planet f} & & &\\
$P_f$ [d] & $12.028_{-0.009}^{+0.009}$&  $12.032_{-0.008}^{+0.007}$   & [12.008,12.049] \\ 
$K_f$ [m/s] & $1.63_{-0.34}^{+0.35}$&  $1.76_{-0.28}^{+0.28}$    & [0.94,2.55] \\ 
$\sqrt{e_f}\cos{\omega_f}$ & $0.006_{-0.13}^{+0.15}$ &  $0.009_{-0.146}^{+0.159}$    & [-0.376,0.429] \\ 
$\sqrt{e_f}\sin{\omega_f}$  & $0.004_{-0.14}^{+0.14}$&  $0.008_{-0.152}^{+0.146}$    & [-0.443,0.444] \\ 
$m_f \sin{i_f}$ [\Mearth] & $6.12_{-1.35}^{+1.33}$ &  $6.64_{-1.05}^{+1.05}$  & [3.52,9.64] \\ 
\hline 
\textit{Planet g} & & &\\
$P_g$ [d] & $17.42_{-0.02}^{+0.03}$&  $17.44_{-0.02}^{+0.01}$   & [17.385,17.478] \\ 
$K_g$ [m/s] & $1.63_{-0.40}^{+0.39}$&  $1.72_{-0.28}^{+0.29}$    & [0.89,2.52] \\ 
$\sqrt{e_g}\cos{\omega_g}$  & $-0.002_{-0.15}^{+0.14}$&  $-0.02_{-0.16}^{+0.15}$    & [-0.457,0.391] \\ 
$\sqrt{e_g}\sin{\omega_g}$  & $-0.01_{-0.15}^{+0.14}$&  $-0.04_{-0.16}^{+0.15}$    & [-0.519,0.383] \\ 
$m_g \sin{i_g}$ [\Mearth]& $6.91 _{-1.76}^{+1.73}$ &  $7.34_{-1.22}^{+1.21}$   & [3.83,10.78] \\ 
\hline 
\end{tabular}
\label{table_params_HD158259}
\end{table*}

\begin{landscape}

\begin{table*}
\centering
\begin{threeparttable}
\caption{Temperatures and pressure measurements on the SOPHIE spectrograph since 2012}    
%\begin{center}
\begin{tabular}{cccccccccccccc}
\hline
Date  & \multicolumn{11}{c}{Temperatures} & \multicolumn{2}{c}{Pressures}\\ & container  & grating & BenchW & BenchE & Cryostat & Coude & ferrule & shutter & spectro & obs room & electronic room & atmosphere & tank\\  

2012-01-01 12:04:25&20.302&19.83&20.156&20.158&19.354&18.571&20.127&20.035&19.777&24.121&20.373&940.601&1002.14\\
2012-01-01 12:10:25&20.307&19.83&20.156&20.158&19.339&18.599&20.127&20.035&19.779&25.108&20.016&940.589&1002.14\\
2012-01-01 12:16:25&20.306&19.83&20.157&20.159&19.337&18.624&20.126&20.035&19.78&24.028&19.82&940.539&1002.14\\
2012-01-01 12:22:25&20.303&19.83&20.157&20.158&19.334&18.649&20.127&20.035&19.779&25.415&19.995&940.485&1002.14\\
2012-01-01 12:28:25&20.304&19.83&20.158&20.158&19.342&18.674&20.126&20.037&19.78&24.135&20.22&940.43&1002.14\\
...&...&...&...&...&...&...&...&...&...&...&...&...&...\\
...&...&...&...&...&...&...&...&...&...&...&...&...&...\\
...&...&...&...&...&...&...&...&...&...&...&...&...&...\\
...&...&...&...&...&...&...&...&...&...&...&...&...&...\\
2022-05-03 22:30:00&20.459&20.20&20.513&20.588&19.062&20.429&20.485&20.417&20.104&24.737&20.098&937.646&1005.51\\
2022-05-03 23:30:00&20.459&20.20&20.515&20.588&19.061&20.427&20.485&20.419&20.103&24.742&20.129&937.727&1005.51\\

\hline

\hline

\end{tabular}

\label{table_housekeeping}
\begin{tablenotes}
      \small

        \item  Temperatures are in degrees and pressure in mbar. Full table available online.
      
    \end{tablenotes}
\end{threeparttable}
\end{table*}

\end{landscape}

%%%%%%%%%%%%%%%%%%%%%%%%%%%%%%%%%%%%%%%%%%%%%%%%%%

%\end{appendix}

\end{document}